\documentclass[%
 reprint,
showpacs,
preprintnumbers,
 amsmath,amssymb,
 aps,
prstab,
]{revtex4-1}

\usepackage{graphicx}
\usepackage{epsfig}
\usepackage{natbib}

\usepackage{dcolumn}
\usepackage{bm}

\usepackage{psfrag}

\def\tdot#1{\stackrel{...}{#1}}

\begin{document}

\preprint{ }

\title{Modeling Longitudinal Oscillations of Bunched Beams in Synchrotrons}

\author{Harald~Klingbeil} 
\author{Dieter~Lens}%
 \altaffiliation{Technische Universit\"{a}t Darmstadt, 64289~Darmstadt, Germany}
 \email{dlens@rtr.tu-darmstadt.de}
\author{Monika~Mehler}
\author{Bernhard~Zipfel}
\affiliation{%
 GSI Helmholtzzentrum f\"{u}r Schwerionenforschung GmbH, Planckstra{\ss}e 1, D-64291 Darmstadt, Germany 
}%




\date{\today}

\begin{abstract}
Longitudinal oscillations of bunched beams in synchrotrons have been analyzed 
by accelerator physicists for decades, 
and a closed theory is well-known \cite{Sacherer1973}. 
The first modes of oscillation are the coherent dipole mode, quadrupole mode, 
and sextupole mode. Of course, these modes of oscillation are included in
the general theory, but for developing RF control systems, it is useful to 
work with simplified models. 
Therefore, several specific models are analyzed in the paper at hand. 
They are useful for the design of closed-loop control systems 
in order to reach an optimum performance with respect to damping the 
different modes of oscillation. This is shown by the comparison of measurement 
and simulation results for a specific closed-loop control system. 
\end{abstract}

                              
\maketitle


\section{Introduction}
%
%
%
%
According to the standard theory, longitudinal bunch oscillations
are characterized by different mode numbers. The mode number 
$m \in\{1,2,3,... \}$ describes the shape of an individual bunch in phase space
whereas the mode number $n \in \{0,1,...,M-1 \}$ describes the phase relation 
between the oscillation of $M$ individual bunches in case there is coupling 
between the bunches (e.g. $n=0$ specifies the case that all bunches are 
oscillating in-phase).

In the paper at hand, only single-bunch oscillations are investigated. 

In longitudinal phase space, the within-bunch modes $m$ may be visualized by 
a symmetrical polygone with $m$ rounded corners. This symmetry 
of course requires appropriate scaling of the phase space coordinates.

If the bunch is located in the linear region of the bucket, i.e. if the 
revolution frequency in phase space approximately equals the synchrotron 
frequency $f_S=1/T_S$ in the bucket center, 
it is obvious that the phase space distribution will be 
the same one as the initial one after the time $T_S/m$. Therefore, it is clear 
that the beam signal (projection of phase space ensemble onto the time axis) 
will contain spectral lines at $m f_S$. Since the same bunch re-appears in a 
ring accelerator after the revolution time $T_R=1/f_R$, 
spectral lines are observed at \citep{Pedersen1977}
\[
f=p M f_R \pm m f_S \mbox{\qquad for } p \in \{0,1,2,\dots \}. 
\]
Therefore, the existence of a spectral line with a specific value $m$ may 
be used to define the mode of oscillation. Using this definition, one 
concludes that a quadrupole mode is present if a spectral line fitting to 
$m=2$ is observed.    

Considering only single-bunch oscillations is not the only simplification 
that is made in this paper. 
Many other effects are not taken into account which are essential
for high-current beam acceleration: 
\begin{itemize}
\item No real beam instabilities are considered. This means that no beam 
impedances are taken into account, and no beam loading is present. 
\item No space charge effects are considered. 
\item No coupling between longitudinal and transverse beam dynamics is assumed.
\end{itemize} 
The exclusion of these effects is usually not relevant for the first design 
of a closed-loop control system. 
Another reason for these simplifications is that in spite of them 
some effects occur which deviate from explanations that 
can be found in literature. 
Therefore, it is clear that one has to be even more careful when 
observations are generalized by adding further physical phenomena 
of practical importance.

In the following sections, several models for longitudinal single-bunch 
oscillations will be presented that describe the phenomenon from different 
points of view. Afterwards, a specific closed-loop control system for damping 
both, dipole and quadrupole oscillations is presented. 
For this system, measurement results are compared with simulation results.    

\section{Models}
\label{models}
In the following, different models for describing longitudinal bunch oscillations
are presented. 

\subsection{Mode Definition by Bunch Shape in Phase Space} 
\label{normal_modes}

\begin{figure}
\centering
\epsfig{file=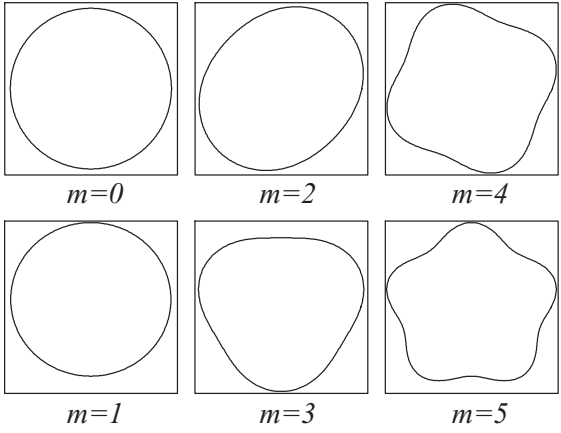,width=3.0in}
\caption{Phase space contour of longitudinal oscillation modes}
\label{figlongoscmodes}
\end{figure}

Assuming that the phase space is scaled in such a way that a matched bunch
in the linear region of the bucket is a circle, the $m$-th mode may 
be described by the formula 
\begin{equation}
r=1+ \epsilon_B \; sin(m \; \varphi).
\label{eqnmodes1}
\end{equation}
Here, $(r,\varphi)$ are the polar coordinates of the bunch contour. 
It is obvious that an unperturbed bunch is obtained for $m=0$.  
 
Fig.~\ref{figlongoscmodes} shows the bunch contours for $\epsilon_B=0.1$. 
 
\subsection{Simplified Linear Differential Equations}

We assume that the RF voltage is modulated according to
\begin{equation}
 u(t) = \hat{u}_0(t) (1+\epsilon(t)) \; 
 sin(\varphi(t) - \Delta \varphi_{gap}(t))
 \label{eqngapvoltage1}
\end{equation}
where 
\[
\varphi(t)=\int \omega_{RF}(t) \; dt. 
\]
By definition, the reference particle arrives at the accelerating gap when 
$\varphi(t)=\varphi_R(t)+2 \pi k$ is
valid (the integer $k$ denotes the bunch repetition number); 
for a non-synchronous particle, the arrival time is defined by
$\varphi(t)=\varphi_R(t)+2 \pi k+\Delta \varphi(t)$. 
The magnetic field $B$ in the bending dipoles 
and the quantities $\hat u_0$, $\omega_{RF}$ and $\varphi_R$ 
are chosen in such a way that the reference particle follows the reference path. 
All these quantities vary slowly with time in comparison with the 
synchrotron oscillation. The modulation functions $\epsilon(t)$ and 
$\Delta \varphi_{gap}(t)$, however, may vary faster. 

Thus, the nonlinear differential equations are
\begin{eqnarray}
\Delta \dot{\varphi} &=& \frac{2 \pi h \eta_R}{T_R \beta_R^2 W_R} \Delta W \label{eqnphi1}\\
\Delta \dot{W} &=& \frac{Q\hat{u}_0}{T_R}[(1+\epsilon) \cdot 
sin (\varphi_R + \Delta \varphi - \Delta \varphi_{gap}) - \nonumber\\
&-& 
sin\; \varphi_R]. \label{eqnw1} 
\end{eqnarray}
Here, $h$ denotes the harmonic number, $T_R$ the revolution time, 
$W_R=m_0 c_0^2 \gamma_R$ the total energy, and  
$\beta_R$, $\gamma_R$ are the 
relativistic Lorentz factors of the reference
particle. With the transition gamma $\gamma_T$, $\eta_R=1/\gamma_T^2-1/\gamma_R^2$ is the
phase slip factor. $Q$ is the charge of one single particle. 
$\Delta W$ and $\Delta \varphi$ are the energy and phase deviations 
of a non-synchronous particle
with respect to the synchronous reference particle (cf.~\citep{lee}). 

For small values of $|\Delta \varphi - \Delta \varphi_{gap}| \ll 1$ we have
\[
\Delta \dot{W} \approx \frac{Q\hat{u}_0 cos\ \varphi_R}{T_R} (1+\epsilon) 
\left( \Delta \varphi - \Delta \varphi_{gap}
+ \frac{\epsilon \; tan \; \varphi_R }{1+\epsilon} \right).
\]
Defining 
\begin{equation}
\Delta \tilde \varphi_{gap} = \Delta \varphi_{gap} - \frac{\epsilon}{1+\epsilon} tan \; \varphi_R
\label{gleqnphigap1}
\end{equation}
leads to
\begin{equation}
\Delta \dot{W} \approx \frac{Q\hat{u}_0 cos\; \varphi_R}{T_R} (1+\epsilon)  
( \Delta \varphi - \Delta \tilde \varphi_{gap}).
\label{eqnw2}
\end{equation}
By a combination of equations (\ref{eqnphi1}) and (\ref{eqnw2}), we get 
\[
\Delta{\ddot \varphi}=\frac{2 \pi h \eta_R Q \hat u_0 cos \; \varphi_R}
{T_R^2 \beta_R^2 W_R} (1+\epsilon)(\Delta \varphi-\Delta \tilde \varphi_{gap}).
\]
The synchrotron frequency is defined by 
\[
\omega_S=2 \pi f_S=
\sqrt{\frac{2 \pi h Q \hat u_0 (-\eta_R \; cos \; \varphi_R)}
{T_R^2 \beta_R^2 W_R}}
\]
which yields 
\begin{equation}
\Delta{\ddot \varphi}+\omega_S^2 (1+\epsilon) \Delta \varphi=
\omega_S^2 (1+\epsilon) \Delta \tilde \varphi_{gap}.
\label{glphaseosc1}
\end{equation}
By using the new variables 
\[
x=\Delta \varphi \mbox{\qquad}
y=C \Delta{\dot \varphi} 
\]
we find: 
\begin{eqnarray*}
\dot x &=& \frac{1}{C} y\\
\dot y &=& -C \omega_S^2 (1+\epsilon)(x-\Delta \tilde \varphi_{gap})
\end{eqnarray*}
For vanishing excitations with $\epsilon=0$ and $\Delta \tilde \varphi_{gap}=0$, we 
now require the trajectories to be circles: 
\[
x=cos(\omega_S t) \Rightarrow 
y=C \dot x=-C \omega_S sin(\omega_S t) \Rightarrow C=-\frac{1}{\omega_S}
\]
Thus, we obtain (note that $C$ and $\omega_S$ vary slowly --- therefore we
neglect the time derivative): 
\begin{eqnarray}
\dot x &=& -\omega_S y \label{gltracking1}\\
\dot y &=& \omega_S (1+\epsilon)(x-\Delta \tilde \varphi_{gap}) \label{gltracking2}
\end{eqnarray}

\subsection{Behavior of Particle Bunches}

Whereas equations (\ref{gltracking1}) and (\ref{gltracking2}) are valid for 
individual particles, we now consider bunches with $N$ particles. 

\subsubsection{Phase Oscillations}

For the mean values, we find: 
\[
\bar x=\frac{1}{N} \sum_{k=1}^N x_k \mbox{\qquad}
\bar y=\frac{1}{N} \sum_{k=1}^N y_k
\]
\begin{equation}
\dot{\bar x}=\frac{1}{N} \sum_{k=1}^N \dot x_k= 
-\omega_S \frac{1}{N} \sum_{k=1}^N y_k=
-\omega_S \bar y 
\label{eqnxbar1}
\end{equation}
\[
\dot{\bar y}=\frac{1}{N} \sum_{k=1}^N \dot y_k= 
\omega_S (1+\epsilon) \frac{1}{N} \sum_{k=1}^N (x_k-\Delta \tilde \varphi_{gap})
\]
\begin{equation}
\Rightarrow 
\dot{\bar y}=\omega_S (1+\epsilon) (\bar x-\Delta \tilde \varphi_{gap})
\label{eqnybar1}
\end{equation}
\begin{equation}
\ddot{\bar x}=-\omega_S \dot{\bar y}=-\omega_S^2 (1+\epsilon) 
(\bar x-\Delta \tilde \varphi_{gap})
\label{gldglxbar1}
\end{equation}
This equation for the bunch center has the same form as 
equation (\ref{glphaseosc1}) for the individual particles. 

\subsubsection{Amplitude Oscillations}

We define the following quantities: 
\[
a_x=\frac{1}{N} \sum_{k=1}^N x_k^2 \mbox{\qquad}
a_y=\frac{1}{N} \sum_{k=1}^N y_k^2 \mbox{\qquad}
\xi=\frac{1}{N} \sum_{k=1}^N x_k y_k
\]
\[
v_x=\frac{1}{N} \sum_{k=1}^N (x_k-\bar x)^2 
=\frac{1}{N} \sum_{k=1}^N (x_k^2-2 x_k \bar x +\bar x^2) 
=a_x-\bar x^2
\]
\[
v_y=a_y-\bar y^2
\]
Please note that $v_x$ corresponds to the variance of the quantities 
$x_k$ if a division by $N-1$ is used instead of the division by $N$. 
Since we are only interested in large numbers $N$, this difference is 
negligible.  

The quantity $\sqrt{v_x}$ represents the bunch length whereas 
$\sqrt{v_y}$ represents 
the height of the bunch (this will later be analyzed in detail). We get: 
\[
\dot a_x=\frac{1}{N} \sum_{k=1}^N 2 x_k \dot x_k=-2 \omega_S \xi 
\]
\[
\dot a_y=\frac{1}{N} \sum_{k=1}^N 2 y_k \dot y_k=
2 \omega_S (1+\epsilon) \frac{1}{N} \sum_{k=1}^N y_k (x_k-\Delta \tilde \varphi_{gap})
\]
\[
\Rightarrow 
\dot a_y=2 \omega_S (1+\epsilon) (\xi-\bar y \Delta \tilde \varphi_{gap})
\]
\begin{eqnarray*}
\dot \xi&=&\frac{1}{N} \sum_{k=1}^N (\dot x_k y_k +x_k \dot y_k)=\\
&=&-\omega_S a_y +   
\omega_S (1+\epsilon)(a_x-\bar x \Delta \tilde \varphi_{gap})
\end{eqnarray*}
\begin{equation}
\dot v_x=\dot a_x-2 \bar x \dot{\bar x}=-2 \omega_S \xi +2 \omega_S \bar x\bar y 
=-2 \omega_S \alpha
\label{glvxdot1}
\end{equation}
Here we defined $\alpha=\xi-\bar x \bar y$
in order to have the same form in the expressions for $a_x$ and for $v_x$.
We find: 
\begin{eqnarray*}
\dot v_y=\dot a_y-2 \bar y \dot{\bar y}&=&
2 \omega_S (1+\epsilon) (\xi-\bar y \Delta \tilde \varphi_{gap})-\\
&-&2 \omega_S (1+\epsilon) \bar y (\bar x-\Delta \tilde \varphi_{gap})
\end{eqnarray*}
\begin{equation}
\Rightarrow 
\dot v_y=2 \omega_S (1+\epsilon) \alpha 
\label{glvydot1}
\end{equation}
\begin{eqnarray*}
\dot \alpha&=&\dot \xi -\dot{\bar x} \bar y-\bar x \dot{\bar y} =\\
&=&-\omega_S a_y +   
\omega_S (1+\epsilon)(a_x-\bar x \Delta \tilde \varphi_{gap})+\\
&+&\omega_S \bar y^2 
-\omega_S (1+\epsilon) \bar x (\bar x-\Delta \tilde \varphi_{gap})
\end{eqnarray*}
\begin{equation}
\Rightarrow
\dot \alpha=-\omega_S v_y+\omega_S (1+\epsilon) v_x
\label{glalphadot1}
\end{equation}
Now we are able to derive a differential equation for $v_x$, i.e. for 
the bunch length oscillation. 

Combining equations (\ref{glvxdot1}) and (\ref{glvydot1}) yields: 
\begin{equation}
\dot v_y=-(1+\epsilon) \dot v_x
\label{glvydot2}
\end{equation}
We now combine equation (\ref{glvxdot1}) with equation (\ref{glalphadot1}):
\begin{equation}
\ddot v_x=2\omega_S^2 v_y-2\omega_S^2 (1+\epsilon) v_x
\label{eqnddotvx1}
\end{equation}
\[
\Rightarrow 
\tdot v_x=2\omega_S^2 \dot v_y-2\omega_S^2 (1+\epsilon) \dot v_x
-2\omega_S^2 \dot \epsilon v_x
\]
Using equation (\ref{glvydot2}), we finally get: 
\begin{equation}
\tdot v_x=-4\omega_S^2 (1+\epsilon) \dot v_x
-2\omega_S^2 \dot \epsilon v_x
\label{gldgl1}
\end{equation}
Please note that for $\epsilon=0$, the standard differential equation 
\begin{equation}
\ddot v_x+(2\omega_S)^2 v_x=const. 
\label{gllindgl1}
\end{equation}
is obtained which corresponds to an oscillation with the frequency $2 \omega_S$. 
Due to the linearization, an initial quadrupole oscillation will continue 
forever. 

Now we derive the differential equation for $v_y$, i.e. for the amplitude 
oscillation. Equation (\ref{glvydot1}) yields: 
\[
\frac{\dot v_y}{1+\epsilon}=2 \omega_S \alpha 
\]
The time derivative is 
\[
\frac{\ddot v_y (1+\epsilon)-\dot \epsilon \dot v_y}{(1+\epsilon)^2}=
-2\omega_S^2 v_y+2\omega_S^2 (1+\epsilon) v_x
\]
where we used eqn.~(\ref{glalphadot1}) on the right side. 
We divide by $(1+\epsilon)$: 
\[
\frac{\ddot v_y}{(1+\epsilon)^2}-\frac{\dot \epsilon \dot v_y}{(1+\epsilon)^3}
=-2\omega_S^2 \frac{v_y}{1+\epsilon}+2\omega_S^2 v_x
\]
Now another time derivative leads to $\dot v_x$ on the right side such that we
can use eqn.~(\ref{glvydot2}) to eliminate $v_x$ completely. After 
some steps, one obtains: 
\begin{eqnarray}
&&\tdot v_y - \frac{3 \ddot v_y \dot \epsilon}{1+\epsilon}+\dot v_y 
\left(
4 \omega_S^2 (1+\epsilon)-
\frac{\ddot \epsilon}{1+\epsilon}+\frac{3 \dot \epsilon^2}{(1+\epsilon)^2} \right)=
\nonumber\\
&&=2 \omega_S^2 \dot \epsilon v_y 
\label{gldgl2}
\end{eqnarray}

This differential equation for $v_y$ differs from eqn.~(\ref{gldgl1}) for 
$v_x$ only by terms that are of higher-order with respect to $\epsilon$.
Furthermore, the sign of the excitation term $2 \omega_S^2 \dot \epsilon v_y$
is different for $v_x$ and $v_y$ which matches the expectations since the
bunch is short when its amplitude is high whereas the bunch is long when 
its amplitude is small.

Please note that we have not introduced any approximations to derive 
the differential equations (\ref{gldgl1}) and (\ref{gldgl2})
from equations (\ref{gltracking1}) and (\ref{gltracking2}). 

According to \citep{Kamke}, these differential equations have the following 
solution: 
\begin{eqnarray}
v_x=C_{x1} w_{x1}^2+C_{x2} w_{x1} w_{x2}+C_{x3} w_{x2}^2 \label{glvx1}\\
v_y=C_{y1} w_{y1}^2+C_{y2} w_{y1} w_{y2}+C_{y3} w_{y2}^2 \label{glvy1}
\end{eqnarray}
The functions $w_{x1}$ and $w_{x2}$ are the linearly independent solutions of 
\[
\ddot w_x+\omega_S^2 (1+\epsilon) w_x=0, 
\]
whereas the functions $w_{y1}$ and $w_{y2}$ are the linearly independent solutions of 
\[
\ddot w_y-\frac{\dot \epsilon}{1+\epsilon} \dot w_y+\omega_S^2 (1+\epsilon) w_y=0.
\]
In the trivial case $\epsilon=0$, we may choose  
\[
w_{x1}=w_{y1}=cos(\omega_S t) \mbox{,\qquad}
w_{x2}=w_{y2}=sin(\omega_S t)
\]
as a solution. Due to equations (\ref{glvx1}) and (\ref{glvy1}), $v_x$ and $v_y$ 
will oscillate with twice the frequency which is in compliance with 
eqn.~(\ref{gllindgl1}).

In appendix \ref{linearization}, it is shown that the 
following differential equations 
are valid as an approximation for small deviations from the matched bunch 
shape: 
\begin{equation}
\ddot{\bar{x}} + \omega_S^2 \bar{x} = \omega_S^2 \Delta \tilde{\varphi}_{gap}
\label{gllindgl2}
\end{equation}
\begin{equation}
\ddot{v}_x + 4\omega_S^2 (v_x-v_0) = -2 \omega_S^2 v_0 \epsilon.
\label{gllindgl3}
\end{equation}

\subsection{Revolution Time in Phase Space}

The revolution frequency of off-center particles in the nonlinear bucket 
in phase space is given by 
\begin{equation}
f_{S,nonlinear}(\Delta \varphi)=
f_S \frac{\pi}{2 K \left(\sin \frac{\Delta \varphi}{2} \right)}
\label{eqnsyncfreq1}
\end{equation}
(cf.~\citep{lee}). 
Here, $\Delta \varphi$ is the maximum phase deviation of the particle and 
$K$ is the complete elliptic integral of the first kind. 
Due to the longitudinal emittance of the bunch in the nonlinear bucket, 
the quantities $\bar x$ and $v_x$ will not 
oscillate with the frequency $f_S$ and $2 f_S$, respectively, but with reduced 
frequencies $f_{S,eff}$ and $2 f_{S,eff}$.   

\begin{figure}
\centering
\psfrag{Rbunch}[c][c]{$2 \sqrt{v_{x0}}$}
\psfrag{Rtrajectory}[c][c]{$\Delta \varphi_{eff}$}
\epsfig{file=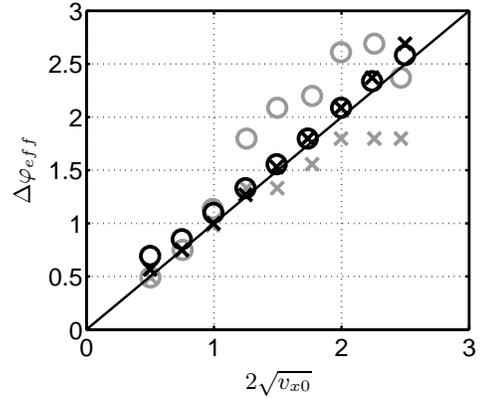,width=2.5in}
\caption{Effective phase deviation $\Delta \varphi_{eff}$ for different 
bunch sizes and different modes of oscillation (\,$\times$: dipole mode, 
$\circ$: quadrupole mode, 
black: homogeneous distribution, grey: Gaussian distribution)}
\label{figeffsyncfreq}
\end{figure}

We now analyzed how $f_{S,eff}$ depends on the size of the bunch. For this 
purpose, the solutions of the linearized ODEs 
(\ref{gllindgl2}) and (\ref{gllindgl3}) 
were compared with a nonlinear tracking simulation for an initial bunch with an
example shape given by
\[
r=r_0 \left[1+0.1 \cos(\varphi)+0.2 \cos(2 \varphi)\right].
\] 
The longitudinal emittance is determined by the parameter $r_0$. 
Due to the mismatch of the bunch, $v_x$ and $v_y$ perform oscillations 
around the average values $v_{x0}$ and $v_{y0}$, respectively. 
Please note that one has $v_{x0} < v_{y0}$ since the nonlinear bucket has the 
shape of an eye instead of a circle. 
For each run of the simulation, the average value $v_{x0}$ 
was determined as a measure for the bunch length. 
Each run of the nonlinear tracking 
simulation also leads to a certain oscillation frequency, and we determined 
$\Delta \varphi$ such that this frequency matches 
$f_{S,nonlinear}(\Delta \varphi)$ for $\bar x$ or      
$2 f_{S,nonlinear}(\Delta \varphi)$ for $v_x$, respectively. As a result, 
the effective phase deviation $\Delta \varphi=:\Delta \varphi_{eff}$ and  
$v_{x0}$ are determined for each simulation run.  

Fig.~\ref{figeffsyncfreq} shows the relation between these two quantities. 
It is obvious that 
the effective synchrotron frequency $f_{S,eff}$ is approximately determined by 
\begin{equation}
f_{S,eff}=f_{S,nonlinear}(\Delta \varphi_{eff})
\label{eqnfseff}
\end{equation}
if 
\begin{equation}
\Delta \varphi_{eff}=2 \sqrt{v_{x0}}
\label{eqndphieff}
\end{equation}
is used.

\subsection{Interpretation of $\bar x$ and $v_x$}

According to the equations
\begin{equation}
\bar x \approx -\varphi_1, \mbox{\qquad}
v_x \approx 2 \; ln \frac{2 \bar i}{A_1}=2 \; ln \frac{A_0}{A_1} 
\label{eqphaseamplitude1}
\end{equation}
the mean value $\bar x$ and the variance $v_x$ may approximately be converted  
into the phase $\varphi_1$ and the amplitude $A_1$ of the fundamental harmonic 
component. $A_0$ is the zeroth Fourier component, and $\bar i$ is the DC 
component of the beam signal.
The approximation (\ref{eqphaseamplitude1}) was derived for an 
elliptical Gaussian bunch with a large number of particles. 
It also holds for a homogeneous distribution. 

We will show now that equation (\ref{eqphaseamplitude1})
is also a good approximation for larger bunches and even if a significant amount 
of filamentation is present. 

\begin{table}
\caption{\label{tabparam}Simulation parameters}
\begin{ruledtabular}
\begin{tabular}{ll}
\textrm{Synchrotron circumference} $l_R$ & $216.72 \; \rm m$ \\ 
\textrm{Transition gamma} $\gamma_T$ & $5.45$ \\ 
\textrm{Ion species} & $\rm {}^{40}Ar^{18+}$ \\ 
\textrm{Kinetic energy} & $11.4 \; \rm MeV/u$\\
\textrm{RF amplitude} $\hat u_0$ & $10 \; \rm kV$ \\
\textrm{Harmonic number} $h$ & 8\\
\textrm{Synchrotron frequency} $f_S$& $3312 \; \rm Hz$ \\ 
\textrm{Revolution time} $T_R$ & $4.66 \; \rm \mu s$\\
\multicolumn{2}{c}{
\sl Parameters resulting from the initial conditions:
} \\
$v_{x0}$ & $0.773$\\
$\Delta \varphi_{eff}=2 \sqrt{v_{x0}}$ & $1.7584$ \\
$f_{S,eff}/f_S$ & $0.8087$ 
\end{tabular}
\end{ruledtabular}
\end{table}

For the parameters shown in Table \ref{tabparam}, solutions for the following 
models were generated numerically: 
\begin{enumerate} 
\item A nonlinear particle tracking simulation. The parameters  
$\bar x$ and $v_x$ are calculated. 
\item An FFT analysis of the previous result was made which leads to the 
phase $\varphi_1$, the amplitude $A_1$ and the DC component $A_0/2=\bar i$. 
Based on these parameters, an approximation for $\bar x$ and $v_x$ is 
calculated using eqn.~(\ref{eqphaseamplitude1}).  
\item Solution of the nonlinear ordinary differential equations (ODE)
(\ref{gldglxbar1}) and (\ref{gldgl1}).
\item Solution of the linearized ODE 
(\ref{gllindgl2}) and (\ref{gllindgl3}). 
\end{enumerate}

For the last two models 3 and 4, the effective synchrotron frequency $f_{S,eff}$
was used instead of $f_S$. 

\begin{figure}
\centering
\psfrag{turns}[c][c]{$t/T_R$}
\psfrag{m1x}{$\bar x$}
\psfrag{m2x}{$v_x$}
\epsfig{file=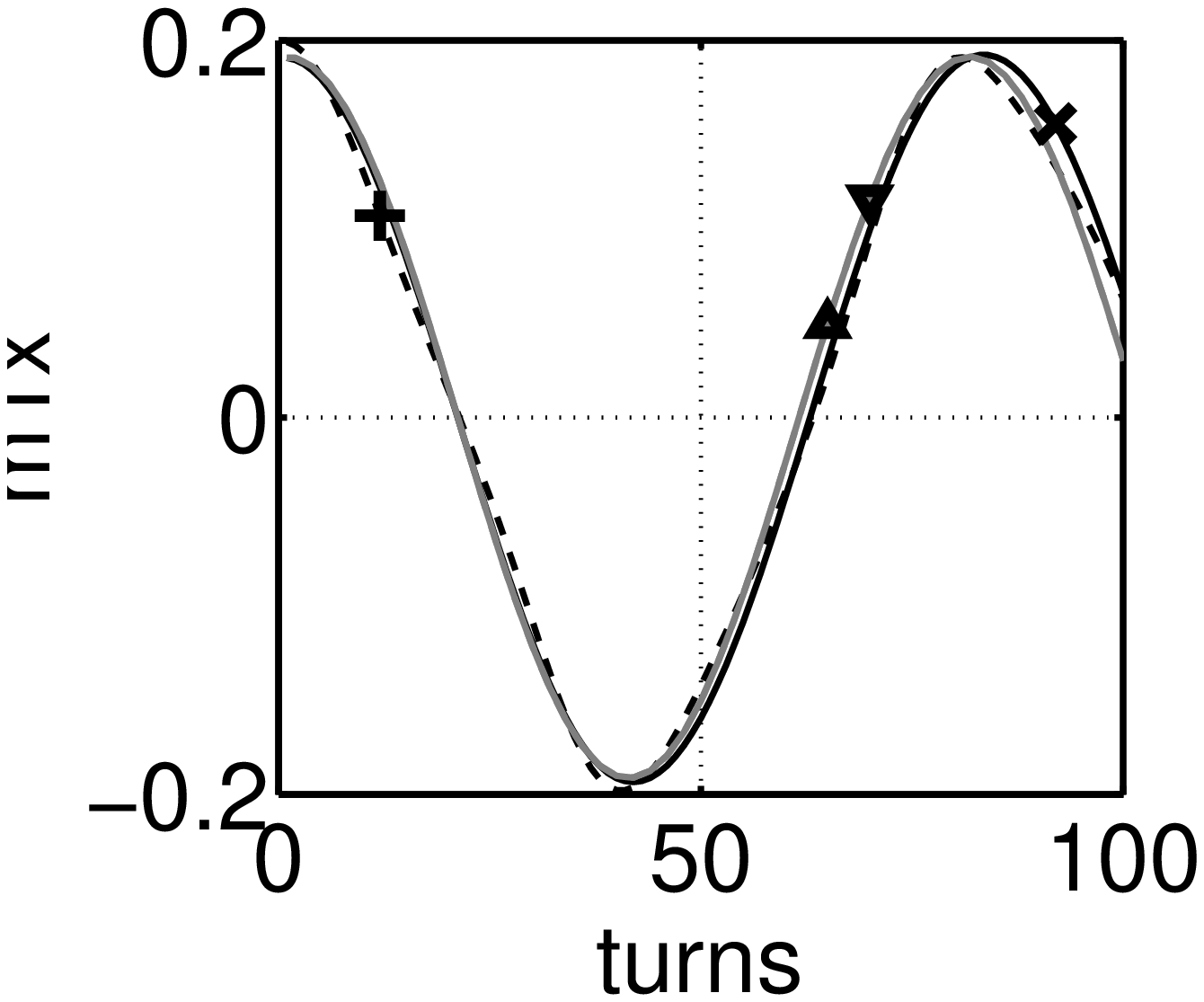,width=1.5in}
\epsfig{file=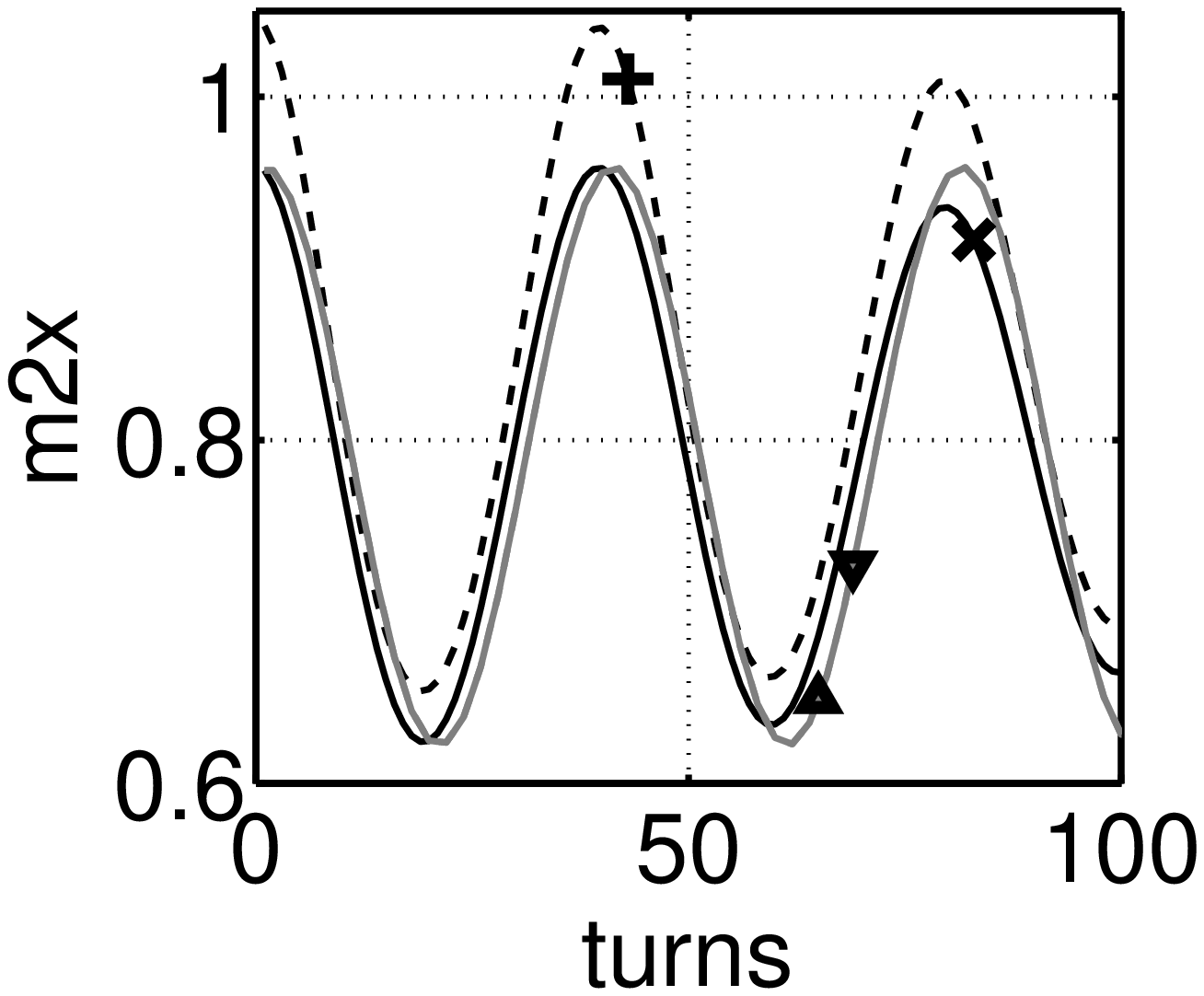,width=1.5in}
\caption{Mean value $\bar x$ and variance $v_x$ versus time
for $\omega_{S,eff}=0.8 \; \omega_S$. The initial 
bunch is a homogeneous distribution according to 
$r=1.6 \left[1+0.1 \cos(\varphi)+0.2 \cos(2 \varphi)\right]$.
No excitation is applied
( $\times$:~nonlinear tracking simulation,  
$+$: result of the FFT using eqn.~(\ref{eqphaseamplitude1}), 
$\triangle$: nonlinear ODE, $\nabla$: linearized ODE).}
\label{figxvxa}
\end{figure}

\begin{figure}
\centering
\psfrag{turns}[c][c]{$t/T_R$}
\psfrag{m1x}{$\bar x$}
\psfrag{m2x}{$v_x$}
\epsfig{file=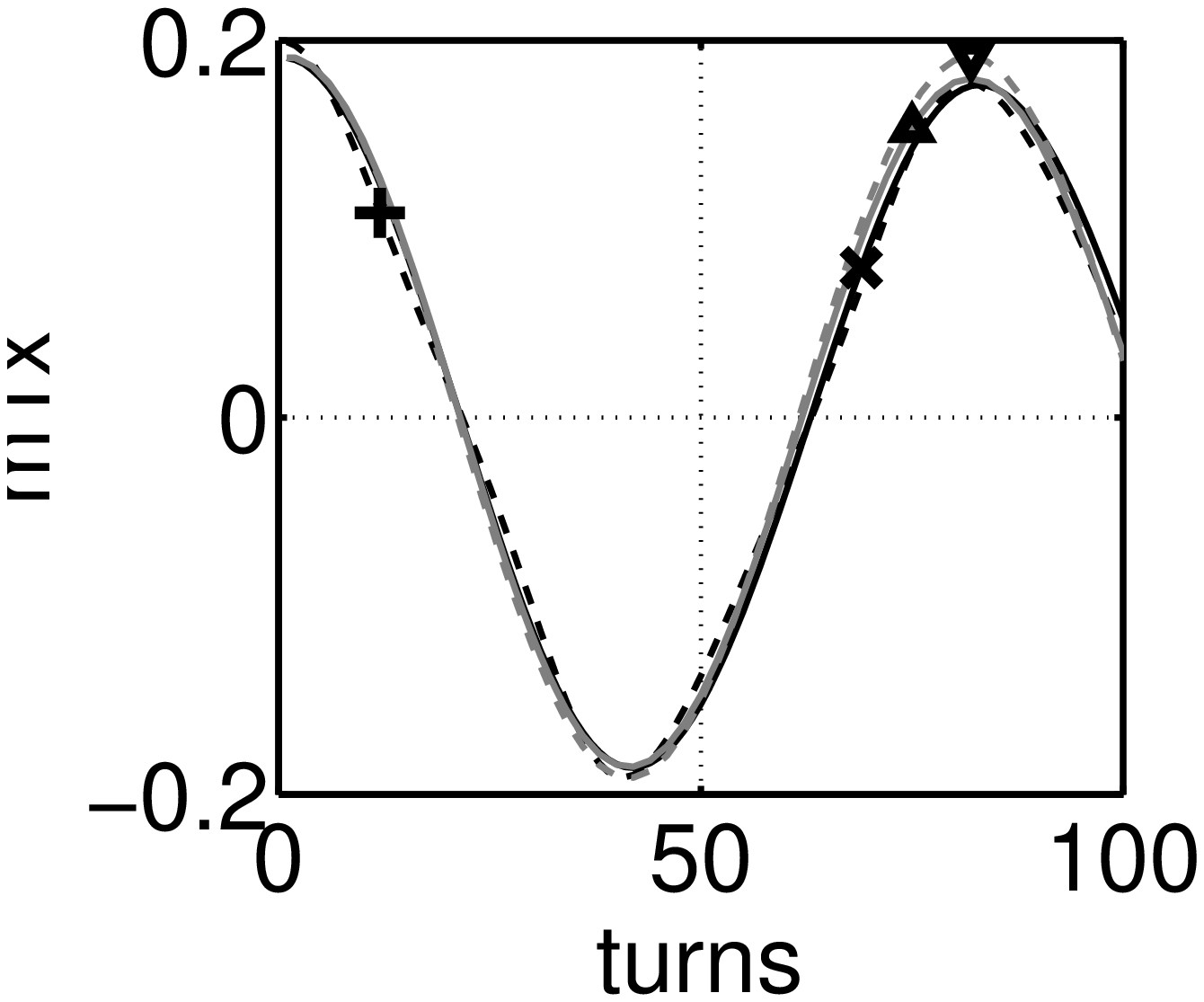,width=1.5in}
\epsfig{file=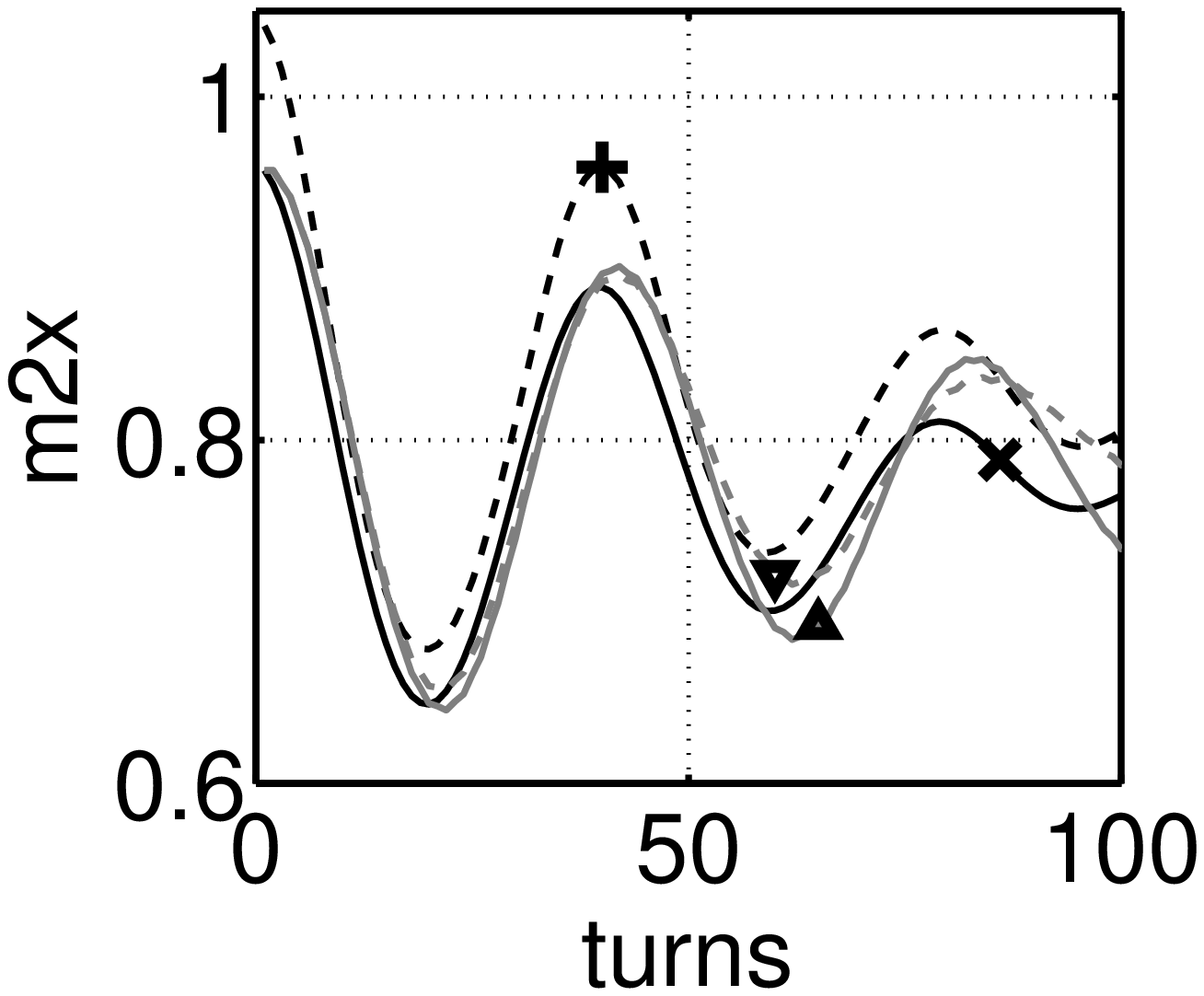,width=1.5in}
\caption{Mean value $\bar x$ and variance $v_x$ versus time
for $\omega_{S,eff}=0.8 \; \omega_S$. The initial 
bunch is a homogeneous distribution according to 
$r=1.6 \left[1+0.1 \cos(\varphi)+0.2 \cos(2 \varphi)\right]$. An excitation
$\epsilon=0.04 \; \sin(2 \omega_{S,eff} t+\pi)$ is applied
( $\times$: nonlinear tracking simulation, 
$+$: result of the FFT using eqn.~(\ref{eqphaseamplitude1}), 
$\triangle$: nonlinear ODE, $\nabla$: linearized ODE).}
\label{figxvxb}
\end{figure}

\begin{figure}
\centering
\psfrag{x}{$x/\rm rad$}
\psfrag{y}{$y$}
\psfrag{Initial}{}
\psfrag{End}{}
\epsfig{file=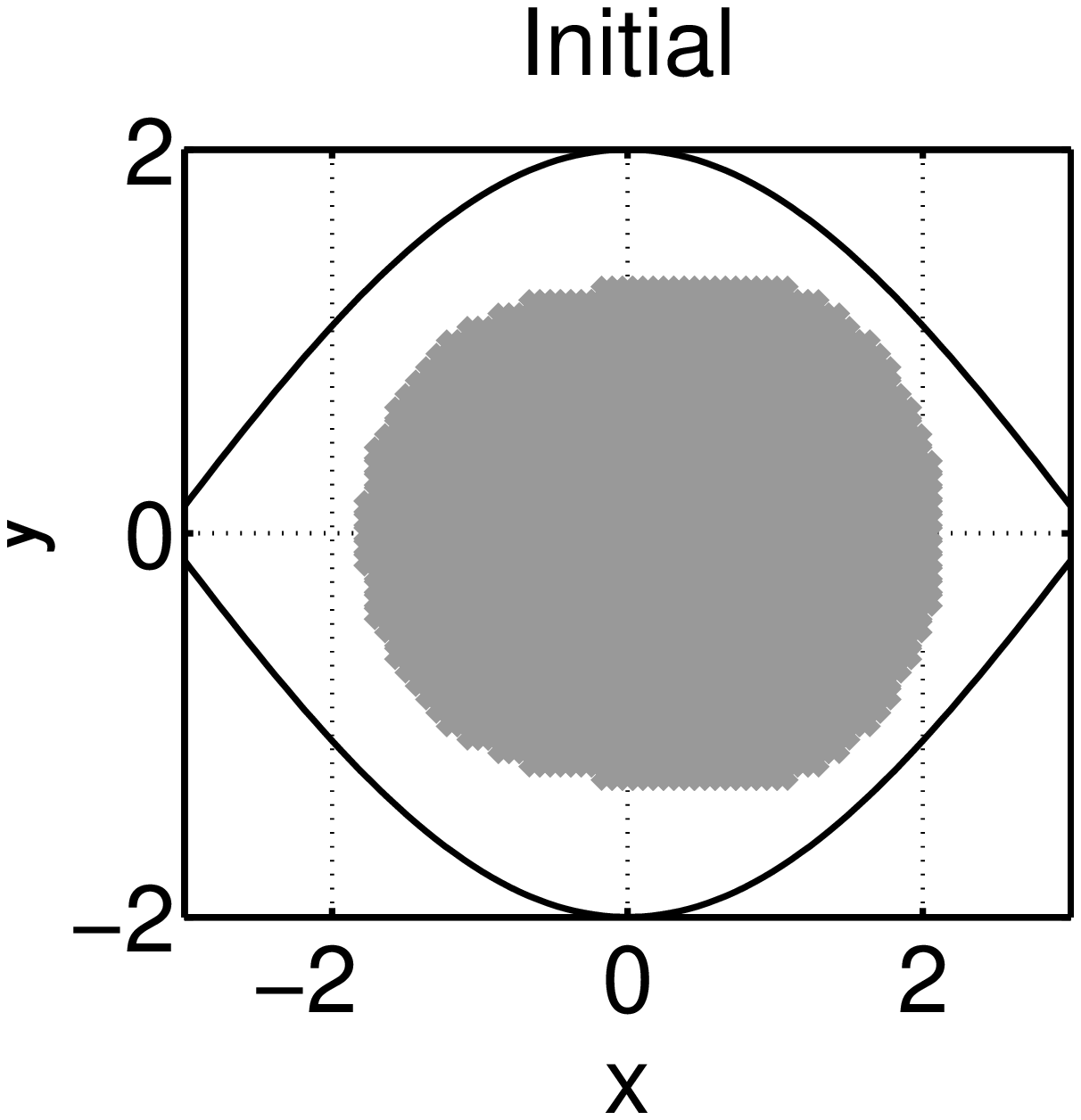,width=1.5in}
\epsfig{file=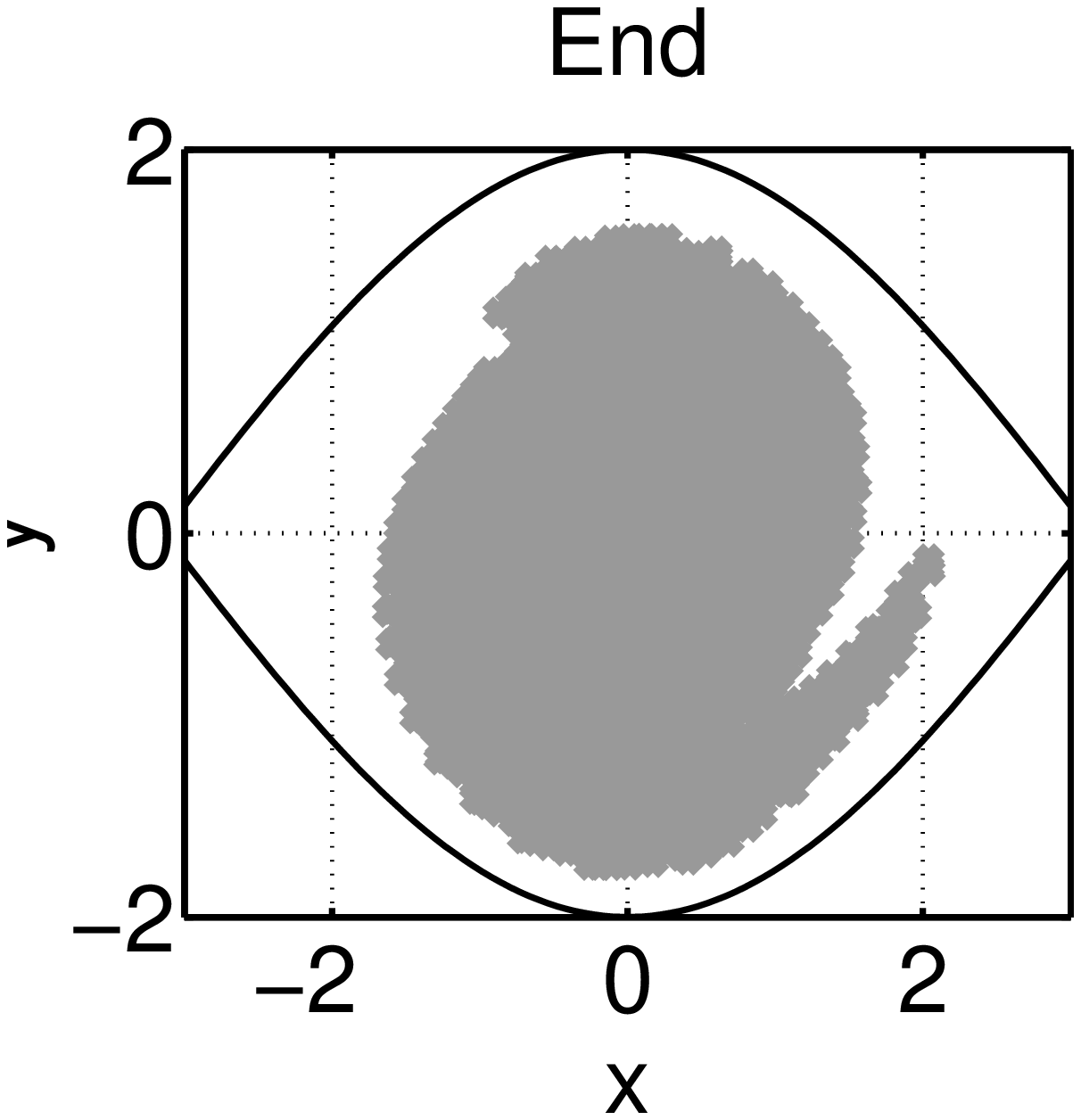,width=1.5in}
\caption{Phase space distribution at the beginning (left diagram) and at the
end (right diagram) of the simulation. The initial 
bunch is a homogeneous distribution according to 
$r=1.6 \left[1+0.1 \cos(\varphi)+0.2 \cos(2 \varphi)\right]$
(mismatch of both, bunch center and bunch length). 
No excitation is applied (same case as in Fig.~\ref{figxvxa}).}
\label{figphasespacea}
\end{figure}

Fig.~\ref{figxvxa} shows the result for a mismatched bunch without 
additional excitation, and Fig.~\ref{figphasespacea} shows the corresponding 
phase space plots. 
Fig.~\ref{figxvxb} shows the result for the same mismatched bunch with 
an additional excitation. 
These examples show the following: 
\begin{itemize} 
\item The formula (\ref{eqphaseamplitude1}) describes the oscillation 
very accurately, but there is a small DC offset which may be relevant for 
small oscillation amplitudes. As we will see later, the 
DC component is usually not of interest for the control loop design.     
\item The models 1 (marker $\times$ in the diagrams) and 2 
(marker $+$ in the diagrams) match very well even for comparatively large 
bunches and for mismatches including filamentation. 
This clearly shows that the quantities $\bar x$ and $v_x$ may be used 
instead of $\varphi_1$ and $A_1$ if the approximation (\ref{eqphaseamplitude1}) 
is used.  
\item The models 1 and 2 include Landau damping since they are based on 
nonlinear tracking equations.  
The models 3 (marker $\triangle$) and 4
(marker $\nabla$) cannot show Landau damping since they are based on linear 
tracking equations.  
\item During the first oscillation period, all models match very well
which indicates that all models may be used for designing feedback systems. 
\item Fig.~\ref{figxvxb} shows that the excitation with $2 f_{S,eff}$ initially leads
to a damping of the amplitude oscillation. The initial damping rates of all 
four models are similar. This is a further confirmation that the models may 
be used for control loop design.  
\end{itemize}

\subsection{Phase and Amplitude Oscillations of the Normal Modes}

We now consider the normal modes introduced in section \ref{normal_modes}. 
If one assumes a homogenous particle distribution inside the bunch contour 
defined by equation (\ref{eqnmodes1}), one may calculate the phase and amplitude 
oscillations analytically. The integrals that have to be solved are the following 
ones: 
\[
M_0=\int_0^{2\pi} \int_0^{r_{max}} r \; dr \; d\varphi
\]
\[
\bar x=\frac{1}{M_0} \int_0^{2\pi} \int_0^{r_{max}} x \; r \; dr \; d\varphi
\]
\[
\bar y=\frac{1}{M_0} \int_0^{2\pi} \int_0^{r_{max}} y \; r \; dr \; d\varphi
\]
\[
a_y=\frac{1}{M_0} \int_0^{2\pi} \int_0^{r_{max}} y^2 \; r \; dr \; d\varphi
\mbox{,\qquad}
v_y=a_y-\bar y^2
\]
Here, the following equations were used: 
\[
r_{max}=1+ \epsilon_B \; sin \left[m \; (\varphi-\varphi_0) \right]
\]
\[
x=r \; cos \; \varphi \mbox{\qquad}
y=r \; sin \; \varphi 
\]
The following results are obtained: 
\[
M_0=\left\{
\begin{array}{ll}
\pi & \mbox{for } m=0\\
\pi \left(1+\frac{1}{2}\epsilon_B^2 \right) & \mbox{for } m > 0
\end{array}
\right.
\]
\[
\bar x=\left\{
\begin{array}{ll}
- \frac{\left(\epsilon_B+\frac{1}{4} \epsilon_B^3 \right) sin \; \varphi_0}
{1+\frac{1}{2} \epsilon_B^2} & \mbox{for } m=1\\
0 & \mbox{for } m \ne 1
\end{array}
\right.
\]
\[
\bar y=\left\{
\begin{array}{ll}
\frac{\left(\epsilon_B+\frac{1}{4}\epsilon_B^3 \right) cos \; \varphi_0}
{1+\frac{1}{2}\epsilon_B^2} & \mbox{for } m=1\\
0 & \mbox{for } m \ne 1
\end{array}
\right.
\]
\[
v_y=\left\{
\begin{array}{ll}
\frac{1}{4} & \mbox{for } m=0\\
\frac{1}{4} \frac{1+ \frac{3}{2} \epsilon_B^2+\frac{7}{8}\epsilon_B^4+\frac{1}{16} \epsilon_B^6
-\frac{1}{2} \epsilon_B^2 \; cos(2\varphi_0)}
{\left(1+\frac{1}{2}\epsilon_B^2 \right)^2} & \mbox{for } m=1\\
\frac{1}{4} \frac{1+ 3 \epsilon_B^2+\frac{3}{8}\epsilon_B^4+\left(2 \epsilon_B
+\frac{3}{2} \epsilon_B^3 \right) sin(2\varphi_0)}
{1+\frac{1}{2}\epsilon_B^2 } & \mbox{for } m=2\\
\frac{1}{4} \frac{1+ 3 \epsilon_B^2+\frac{3}{8}\epsilon_B^4}
{1+\frac{1}{2}\epsilon_B^2 } & \mbox{for } m>2\\
\end{array}
\right.
\]
We see that only the mode $m=1$ shows a phase modulation with the frequency $f_S$
(since $\bar x$ is periodic with respect to $\varphi_0$).
The mode $m=1$ also shows an amplitude modulation with the frequency $2 f_S$
(since $v_y$ is periodic with respect to $2\varphi_0$), but
this is a parasitic effect caused by the fact that the bunch shape is only 
approximately circular. The slight deformation causes the modulation of the order 
$\epsilon_B^2$. Apart from this exception, only the mode $m=2$ shows an 
amplitude modulation with the frequency $2 f_S$ 
of the order $\epsilon_B$. The modes $m>2$ neither show 
a phase modulation nor an amplitude modulation. 

\subsection{Spectrum of Dipole Oscillation}

\begin{figure}
\centering
\epsfig{file=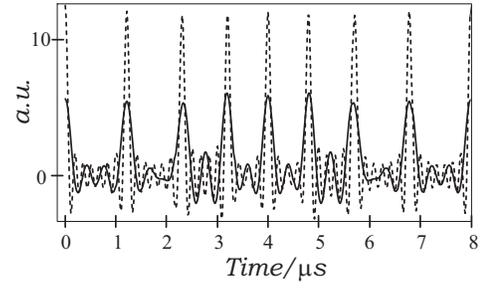,width=2.5in}
\caption{Visualization of the partial sums of the series in eqn.~(\ref{eqndirac1}) 
for $T_R=1 \;\mu s$, $T_S=8 \;\mu s$, $\Delta t=0.3 \; \mu s$
(solid line: summation up to $n=20$, dashed line: summation up to $n=50$)}
\label{figdiracdipole}
\end{figure}

\begin{figure}
\centering
\epsfig{file=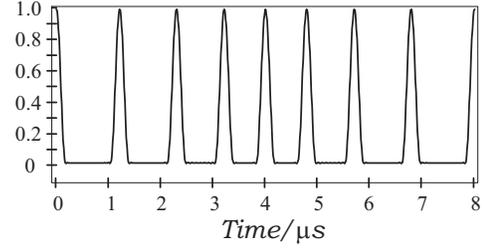,width=2.5in}
\caption{Dipole oscillation of $cos^2$-shaped bunches  
for $T_R=1 \;\mu s$, $T_S=8 \;\mu s$, $\Delta t=0.3 \; \mu s$, $\Omega=18/T_R$
(summation up to $n=100$)}
\label{figbunchdipole}
\end{figure}

In appendix \ref{fourier_series}, we prove the formula 
\begin{eqnarray}
\sum_{k=-\infty}^{+\infty} 
\delta\left(t-k T_R-\Delta t \; sin\left(2 \pi k \frac{T_R}{T_S} \right)\right)=
\nonumber \\
=\frac{1}{T_R}+
\sum_{n=1}^\infty A_n cos(n \omega_S t)
\label{eqndirac1}
\end{eqnarray}
for $\Delta t<T_R/2$ where 
\[
A_n=\frac{2}{T_S} 
\left[1+\sum_{k=1}^{\frac{T_S}{T_R}-1} cos \left(n \omega_S 
\left[k T_R+\Delta t \, sin \frac{2 \pi k T_R}{T_S} \right]
\right)\right]
\]
and $T_S/T_R$ is a positive integer. 

Fig.~\ref{figdiracdipole} shows the graph of two different partial sums of the 
series on the right side of eqn.~(\ref{eqndirac1}) for one period 
$T_S=2\pi/\omega_S$. 
It is obvious that the peaks become the higher 
the more terms are added. One also sees that the pulse density is high in the
middle whereas it is low at the beginning and at the end of the period. 
This is the expected behavior for the dipole oscillation under consideration. 

A real beam signal consists of pulses with finite height and length instead 
of the Dirac pulses. Let us assume that such a single finite pulse centered at 
$t=0$ is given by the function $x(t)$.
If we write the above-mentioned train of Dirac pulses 
as 
\[
h(t)=\sum_{k=-\infty}^{+\infty} \delta(t-T_k),
\]
we get the convolution 
\begin{equation}
y(t)=h(t)*x(t)=\sum_{k=-\infty}^{+\infty} x(t-T_k),
\label{eqnconvol1}
\end{equation}
which corresponds to the desired beam signal. This signal still shows dipole 
oscillations if $T_k$ is implicitly defined by eqn.~(\ref{eqndirac1}). 
Since 
\[
h(t)=\sum_{n=-\infty}^{+\infty} c^{(h)}_n e^{jn\omega_S t}
\]
is given by a Fourier series, the corresponding Fourier transform 
is 
\[
H(j\omega)=2\pi \sum_{n=-\infty}^{+\infty} c^{(h)}_n \delta(\omega-n \omega_S)
\]
where $c^{(h)}_n=\frac{A_n}{2}$. The Fourier transform of eqn.~(\ref{eqnconvol1})
is 
\begin{eqnarray*}
Y(j\omega)&=&H(j\omega) X(j\omega)=\\
&=&
2\pi \sum_{n=-\infty}^{+\infty} c^{(h)}_n X(jn\omega_S)\delta(\omega-n \omega_S).
\end{eqnarray*}
This is again a Fourier series whose Fourier coefficients are given by
\[
c^{(y)}_n=c^{(h)}_n X(j n \omega_S).
\]
For beam pulses that have the form 
\[
x(t)=\left\{
\begin{array}{ll}
cos^2 \left(\frac{\Omega}{2} t \right) & \mbox{\quad for } -\pi\le \Omega t \le\pi\\
0 & \mbox{\quad elsewhere}
\end{array}
\right.
\]
one may --- as a simple but still realistic example ---
derive the Fourier transform 
\[
X(j\omega)=\left\{
\begin{array}{ll}
\frac{\pi}{\Omega} 
\frac{si\left(\pi \frac{\omega}{\Omega}\right)}
{1-\left(\frac{\omega}{\Omega} \right)^2} & \mbox{ for } |\omega| \ne \Omega\\
\frac{\pi}{2\Omega} & \mbox{ for } |\omega| = \Omega
\end{array}
\right. 
\]
An example for such a bunch train with dipole oscillations is shown in 
Fig.~\ref{figbunchdipole}. In this example, we have chosen an unrealistic 
synchrotron period of $T_S=8\; \mu s$ in order to be able to show the result 
in a diagram. 
For a realistic synchrotron period of $T_S=1000\; \mu s$, the Fourier 
coefficients are shown in Table \ref{table_fourier_bunchtrain}. 
Please note that the following condition has to be fulfilled since the bunch  
has to fit into the bucket even if the maximum bunch offset $\Delta t$ occurs:
\[
\frac{\pi}{\Omega}+\Delta t < \frac{T_R}{2}
\]

\begin{table*}
\caption{\label{table_fourier_bunchtrain}Fourier coefficients $a^{(y)}_n=2 c^{(y)}_n$ of a bunch train with 
dipole ocscillations
($T_R=1 \;\mu s$, $T_S=1000 \;\mu s$, $\Delta t=0.3 \; \mu s$, $\Omega=18/T_R$)
}
\begin{ruledtabular}
\begin{tabular}{r|rrrr}
 &        $f_R$ & $2 f_R$ & $3 f_R$ & $4 f_R$ \\ \hline 
$-5 f_S$ & 0.0016819 & 0.0267036 & 0.0545552 & 0.0222028\\
$-4 f_S$ & 0.0087296 & 0.0618662 & 0.0634047 & 0.0012028\\
$-3 f_S$ & 0.0355937 & 0.1048087 & 0.0351619 & -0.0208789\\
$-2 f_S$ & 0.1050056 & 0.1050122 & -0.0260749 & -0.0177420\\
$-1 f_S$ & 0.1875566 & 0.0065220 & -0.0534534 & 0.0114866\\
$+0 f_S$ & 0.0937025 & -0.1014007 & 0.0073288 & 0.0207027\\
$+1 f_S$ & -0.187475 & -0.0061292 & 0.0533306 & -0.0115380\\
$+2 f_S$ & 0.1055102 & 0.1045001 & -0.0263491 & -0.0175559\\
$+3 f_S$ & -0.0361045 & -0.1048387 & -0.0345397 & 0.0208711\\
$+4 f_S$ & 0.0089756 & 0.0622868 & 0.0629741 & 0.0008722\\
$+5 f_S$ & -0.0017600	& -0.0271106 & -0.0545328 & -0.0217486
\end{tabular}
\end{ruledtabular}
\end{table*}

The following points are observed: 
\begin{itemize}
\item The dominant spectral lines do not occur at $p f_R \pm f_S$ in general. 
In our example, this is only valid for $p=1$. 
\item Although an ideal dipole mode was modeled, spectral lines also exist at 
$p f_R \pm 2 f_S$, $p f_R \pm 3 f_S$, etc. This is normally interpreted as 
a quadrupole, sextupole, etc. component in the oscillation. 
\item The sidebands are not symmetric with respect to the harmonics of the 
revolution frequency. 
\end{itemize}

When $\Delta t$ is reduced significantly in this example (i.e. for smaller 
dipole oscillations), the dominant spectral lines are located at  
$p f_R \pm f_S$ as expected, and the symmetry is also improved. The 
pure existence of Fourier components at 
$p f_R \pm 2 f_S$, $p f_R \pm 3 f_S$, etc. remains, however 
(their magnitude is smaller than that at $p f_R \pm f_S$).

It has to be emphasized that the validity of these observations is very general. 
Even though the example was a special one, the formulas for the Fourier series 
are not based on any approximations.

\section{Sextupole Mode Generation} 

In this section, we try to excite a sextupole oscillation by a phase
modulation with 
\[
\Delta \varphi_{gap}=\frac{\pi}{18} \; sin(3 \omega_S t)
\]
which corresponds to a phase deviation of $\pm 10^\circ$. 
The parameters defined in Table \ref{tabparam} are used, and a matched bunch 
($r=1$) is assumed at the beginning. 

\begin{figure}
\centering
\psfrag{nonlinear}[c][c]{}
\psfrag{linear}[c][c]{}
\psfrag{x}{$x$}
\psfrag{y}{$y$}
\epsfig{file=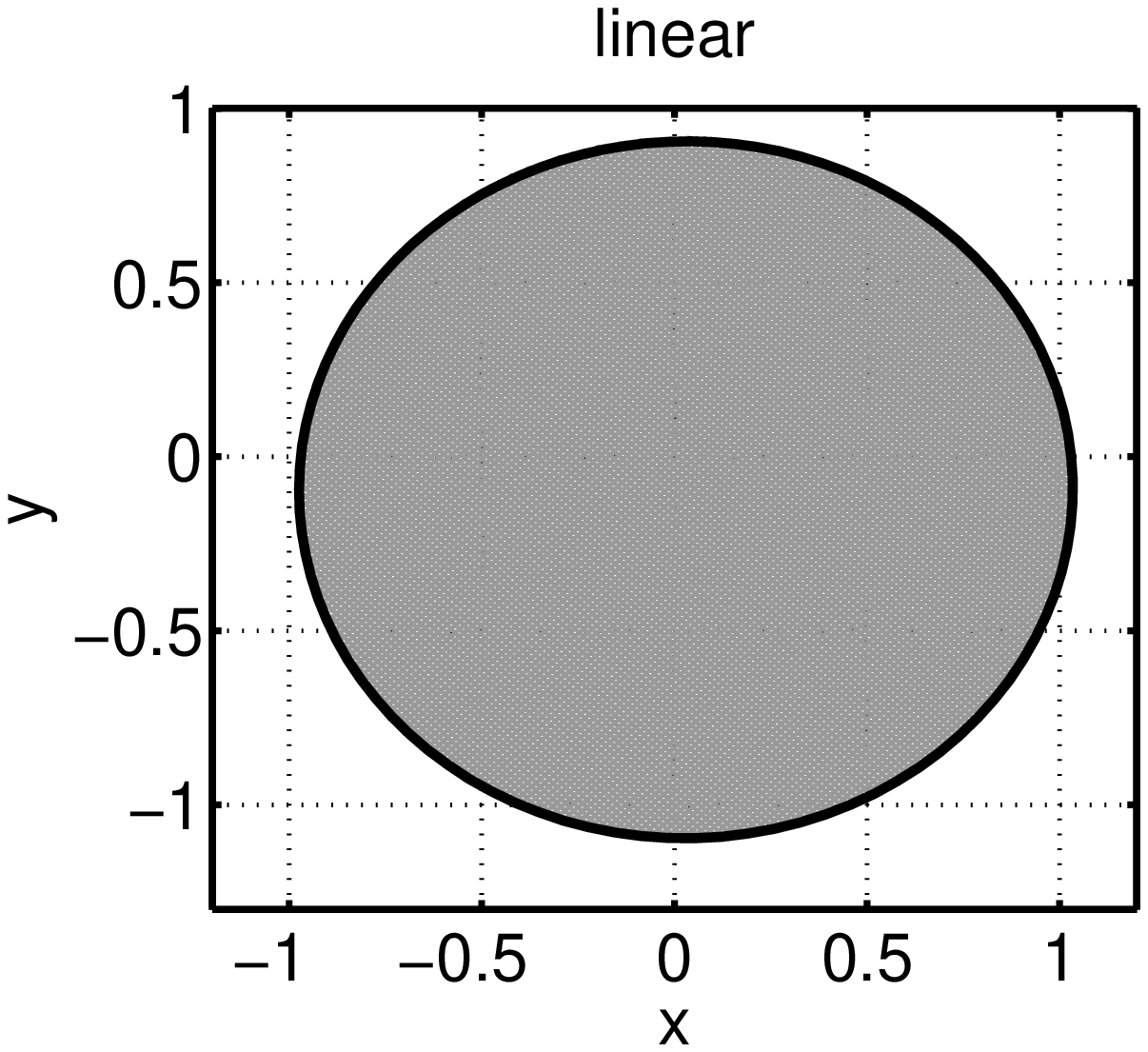,width=1.5in}
\epsfig{file=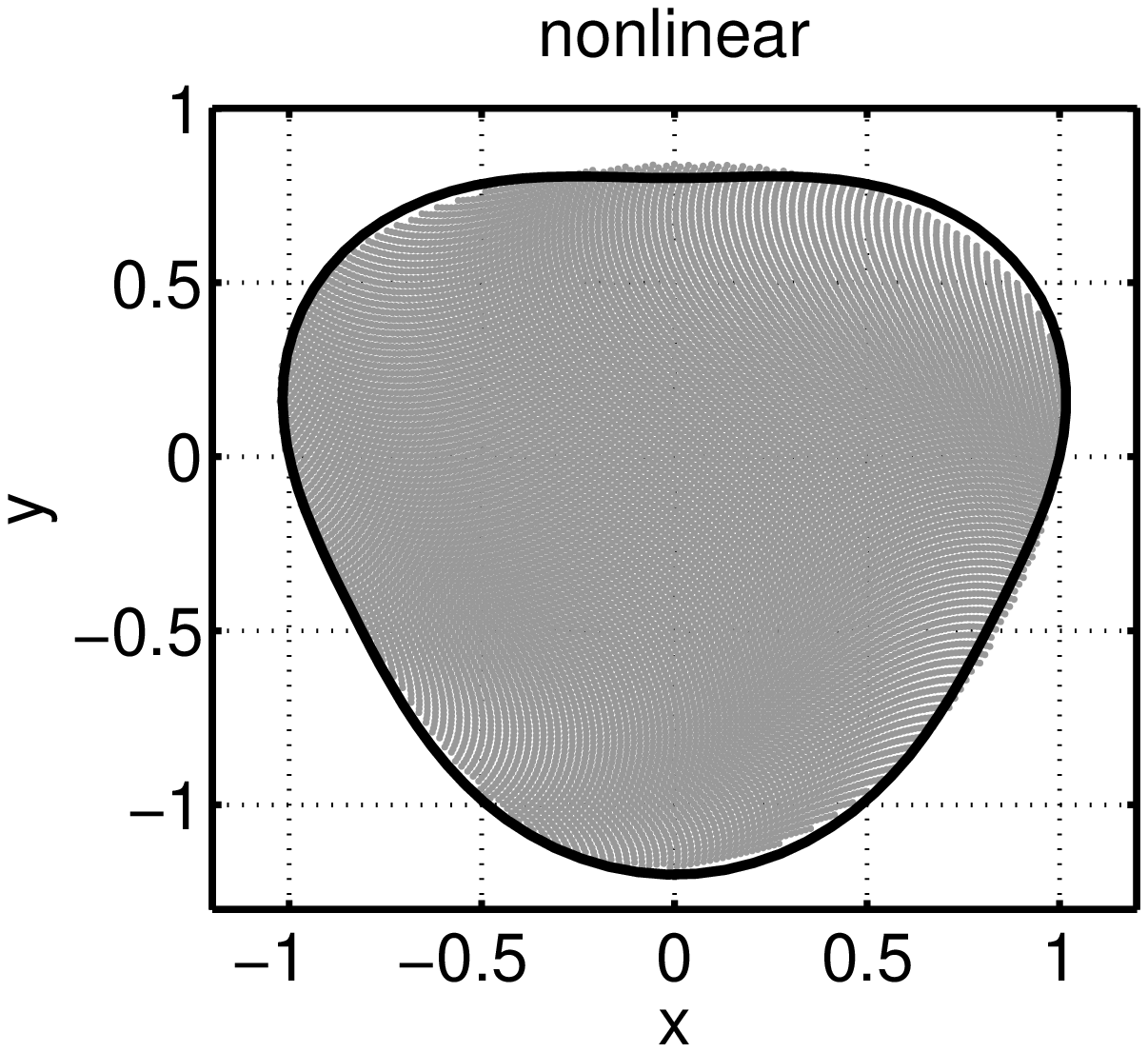,width=1.5in}
\caption{Phase space at $t=1.12 T_S$ for an excitation 
with $3f_S$ (Left diagram: linear tracking equations, 
right diagram: nonlinear tracking equations)}
\label{figsextupole}
\end{figure}

After a time $t=1.12 \; T_S$, the phase space distribution shown in 
Fig.~\ref{figsextupole} is obtained. The left diagram shows that no typical 
sextupole distribution is obtained if linearized tracking equations are used. 
In the right diagram one can see that the original nonlinear tracking equations 
lead to a sextupole distribution whose contour is compliant with equation 
(\ref{eqnmodes1}) for $m=3$. Matching the contour to the particle cloud 
leads to 
\[
r=1+0.1 \; sin(\varphi-\pi)+0.1 \; sin(3 \varphi).
\]
This shows that both, a dipole and a sextupole mode is excited. It has to be 
emphasized that an excitation of the sextupole mode is only possible in the
nonlinear bucket. This was also verified using the program package 
ESME \citep{MacLachlan1997} and 
can furthermore be shown analytically \citep{Gross}.

\section{Application of Models}

Using the models described in section \ref{models}, an example for a specific 
longitudinal damping system is presented in the following. 
The theoretical results are compared with measurement data. 

\subsection{Beam Phase Control and Quadrupole Damping System}

\begin{figure}
\centering
\epsfig{file=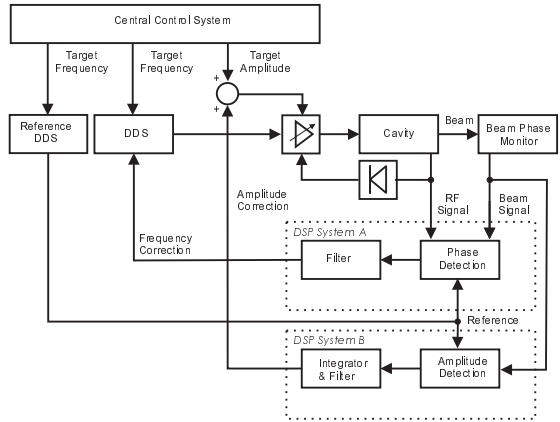,width=3in}
\caption{Block diagram of the beam phase control system and the quadrupole
damping system}
\label{figblockdiagram}
\end{figure}

The models presented before were used to analyze an RF system that 
is displayed in Fig.~\ref{figblockdiagram}. A DDS module provides an RF master 
signal with the frequency provided by the central control system. An amplitude 
control loop which is not considered in the paper at hand makes sure that 
the detected amplitude of the cavity RF signal matches the target amplitude 
provided by the central control system. 

A phase detection algorithm in DSP system A 
compares the phase of the cavity voltage with the phase of the beam signal. 
This phase difference is then processed by a digital bandpass filter with 
$f_S$ belonging to the passband. After applying a proportional gain, the filter 
output is used to modify the DDS frequency. For analyzing the loop 
characteristics, the DDS may be regarded as an integrator with respect to the 
phase. The loop A ensures that coherent 
dipole oscillations with a frequency approximately equal to $f_S$ will be 
damped.

In DSP system B, an amplitude detection algorithm is applied to the beam signal. 
This signal is fed to a digital filter with a passband 
frequency of about $2 f_S$. In contrast to the dipole oscillation damping loop A
mentioned before, there is no intrinsic integrator in loop B. Therefore, an 
additional integrator is implemented in DSP system B. The resulting signal is
used to modify the target amplitude in order to damp quadrupole oscillations.    

Both DSP systems need a reference signal which allows them to detect RF signals
at the relevant frequency. For this purpose a reference DDS is used.  

\subsection{Analytic Model for the Damping System} 

The first natural step to analyze a control loop system like that shown in
Fig.~\ref{figblockdiagram} is to linearize the building blocks in order to 
get transfer functions in the Laplace domain. For example, 
eqn.~(\ref{gllindgl2}) directly leads to the well-known~\citep{Boussard1991}
beam transfer function 
\[
G_1(s)=\frac{\Delta \varphi_B}{\Delta \tilde{\varphi}_{gap}}=\frac{\omega_S^2}{s^2+\omega_S^2}
\] 
where $\Delta \varphi_B=\bar x$ is the beam phase. 
Eqn. (\ref{gllindgl3}) is more difficult to interpret. In a well-designed 
quadrupole damping system, no significant emittance blow-up will occur. Therefore, 
the variance $v_0$ is a constant which is defined by the longitudinal emittance
of the bunches. Hence, by defining 
\[
\Delta v_x=v_x-v_0
\]
we may rewrite eqn.~(\ref{gllindgl3}) in the form 
\[
\Delta \ddot v_x + 4\omega_S^2 \; \Delta v_x = -2 \omega_S^2 v_0 \epsilon.
\]
which corresponds to the transfer function 
\[
G_2(s)=\frac{\Delta v_x}{-2 v_0 \epsilon}=\frac{\omega_S^2}{s^2+4\omega_S^2}.
\]
which is also known from literature \citep{Boussard1991}. 


\begin{figure}
\centering
\psfrag{Dipole}[c][c]{}
\psfrag{Quadrupole}[c][c]{}
\psfrag{chi}{$f_{pass}/f_S$}
\psfrag{K-omega}{$K_{crit}/\omega_S$}
\epsfig{file=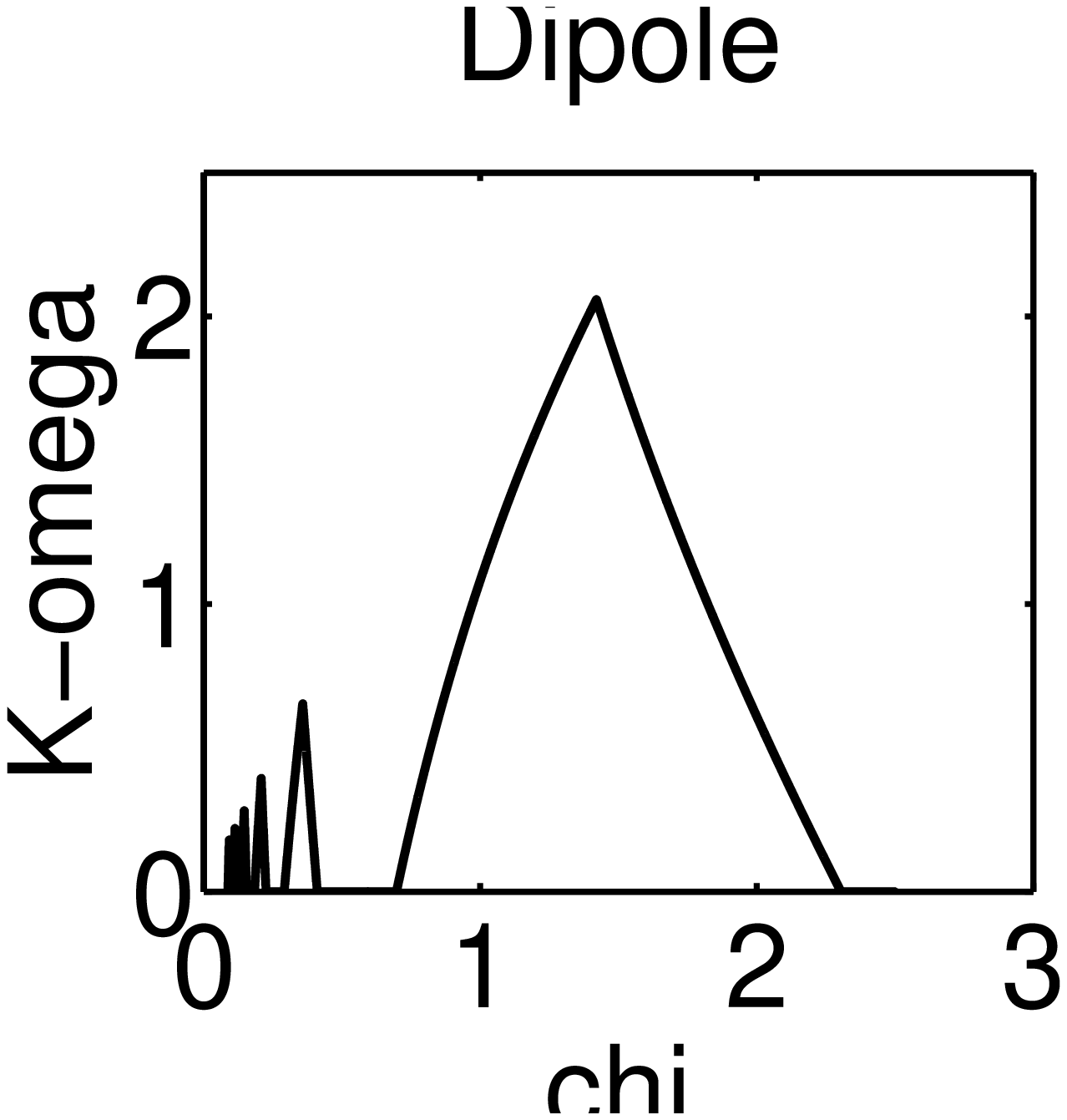,width=1.5in}
\psfrag{chi}{$f_{pass}/2 f_S$}
\epsfig{file=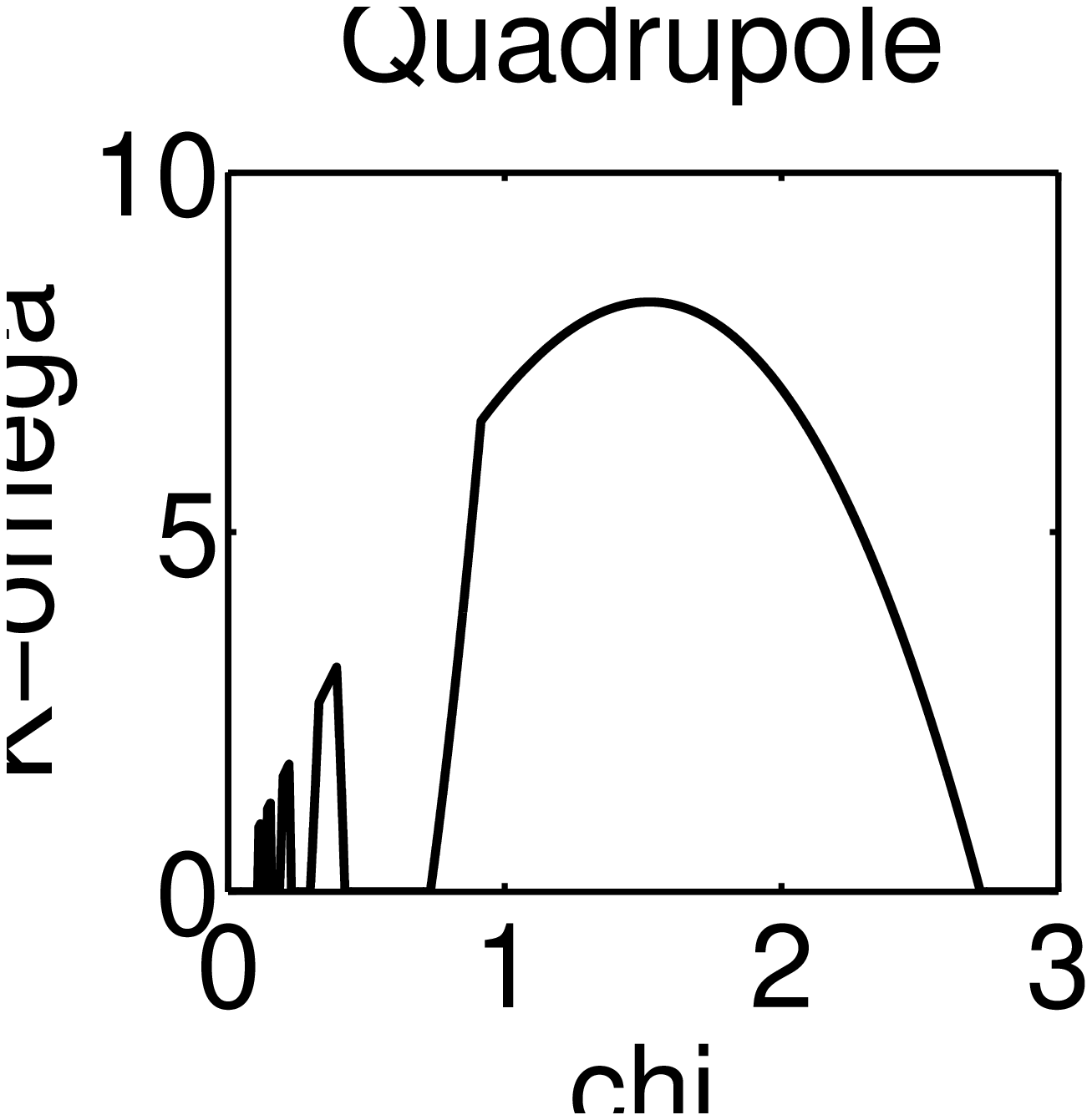,width=1.5in}
\caption{Stability diagram for damping dipole (left diagram) and 
quadrupole modes (right diagram), $T_d=10\; \rm \mu s$, $f_S=3312 \; \rm Hz$}
\label{figstabilty}
\end{figure}

By using the transfer function $G_1(s)$, the beam-phase control loop was
analyzed in \citep{Klingbeil2007}. In an analog way we took $G_2(s)$ to 
analyze the quadrupole damping loop. In both cases, the same type of filter 
specified in \citep{Klingbeil2007} was used. 
Fig.~\ref{figstabilty} shows the region of stability for both, the dipole and 
the quadrupole damping system. Formulas for the first one can be 
found in \citep{Klingbeil2007}, the stability region for the latter one 
is given by
\[
K_{crit}=\min_p K_{p,crit}
\]
where
\[
\frac{K_{p,crit}}{\omega_S}=16 \frac{\xi |1-\xi^2|}{1-\cos \frac{\pi \xi}{\chi}}, 
\]
\[
\xi=\left\{
\begin{array}{ll}
\frac{p-\frac{1}{2}}{\frac{1}{\chi}+4 T_d f_S} & 
\mbox{ for } \xi<1 \mbox{ and } p\in\{1,3,5,\dots\}\\ 
\frac{p+\frac{1}{2}}{\frac{1}{\chi}+4 T_d f_S} & 
\mbox{ for } \xi>1 \mbox{ and } p\in\{1,3,5,\dots\} 
\end{array}
\right.
\]  
and 
\[
\chi=\frac{f_{pass}}{2 f_S}.
\] 
These results were derived in the same way as described in \citep{Klingbeil2007}. 
The stability diagram is in compliance with the measurement results.

\subsection{Measurement Results}
\label{measresults}

\begin{table}
\caption{\label{tabparambeamexp}Beam experiment parameters}
\begin{ruledtabular}
\begin{tabular}{ll}
\textrm{Synchrotron circumference} $l_R$ & $216.72 \; \rm m$ \\ 
\textrm{Transition gamma} $\gamma_T$ & $5.45$ \\ 
\textrm{Ion species} & $\rm {}^{40}Ar^{18+}$ \\ 
\textrm{Kinetic energy} & $11.4 \; \rm MeV/u$\\
\textrm{DC beam current} $\bar i$ & $2 \; \rm mA$\\
\textrm{RF amplitude} $\hat u_0$ & step from $5$ to $10 \; \rm kV$ \\
\textrm{Harmonic number} $h$ & 8\\
\textrm{Synchrotron frequency $f_S$ at $10 \; \rm kV$} & $3312 \; \rm Hz$ \\ 
\textrm{Revolution time} $T_R$ & $4.66 \; \rm \mu s$
\end{tabular}
\end{ruledtabular}
\end{table}

\begin{table}
\caption{\label{tabloopparams}Control loop parameters for measurement and simulation}
\begin{ruledtabular}
\begin{tabular}{lll}
\textrm{Delay time dipole damping system} & $10 \; \rm \mu s$\\ 
\textrm{Delay time quadrupole damping system} & $10 \; \rm \mu s$\\
\textrm{Filter frequency quadrupole damping system} & $9000 \; \rm Hz$\\
\textrm{Filter frequency dipole damping system} & $3500 \; \rm Hz$\\
\textrm{Sampling frequency of FIR filters} & $375.44 \; \rm kHz$
\end{tabular}
\end{ruledtabular}
\end{table}

In order to verify the models presented here, a closed-loop analysis was 
performed, and the results were compared with the measurement results
obtained in a beam experiment. Table \ref{tabparambeamexp} shows the 
beam experiment conditions. In this experiment, a matched beam in 
a stationary bucket generated by an RF voltage of $5 \; \rm kV$ was 
present. By doubling the voltage instantaneously, quadrupole oscillations 
were excited intentionally. Excitations of this magnitude will never occur 
in practice --- they were only used to show the functionality of the system 
and the validity of the theory. 

\begin{figure}
\centering
\epsfig{file=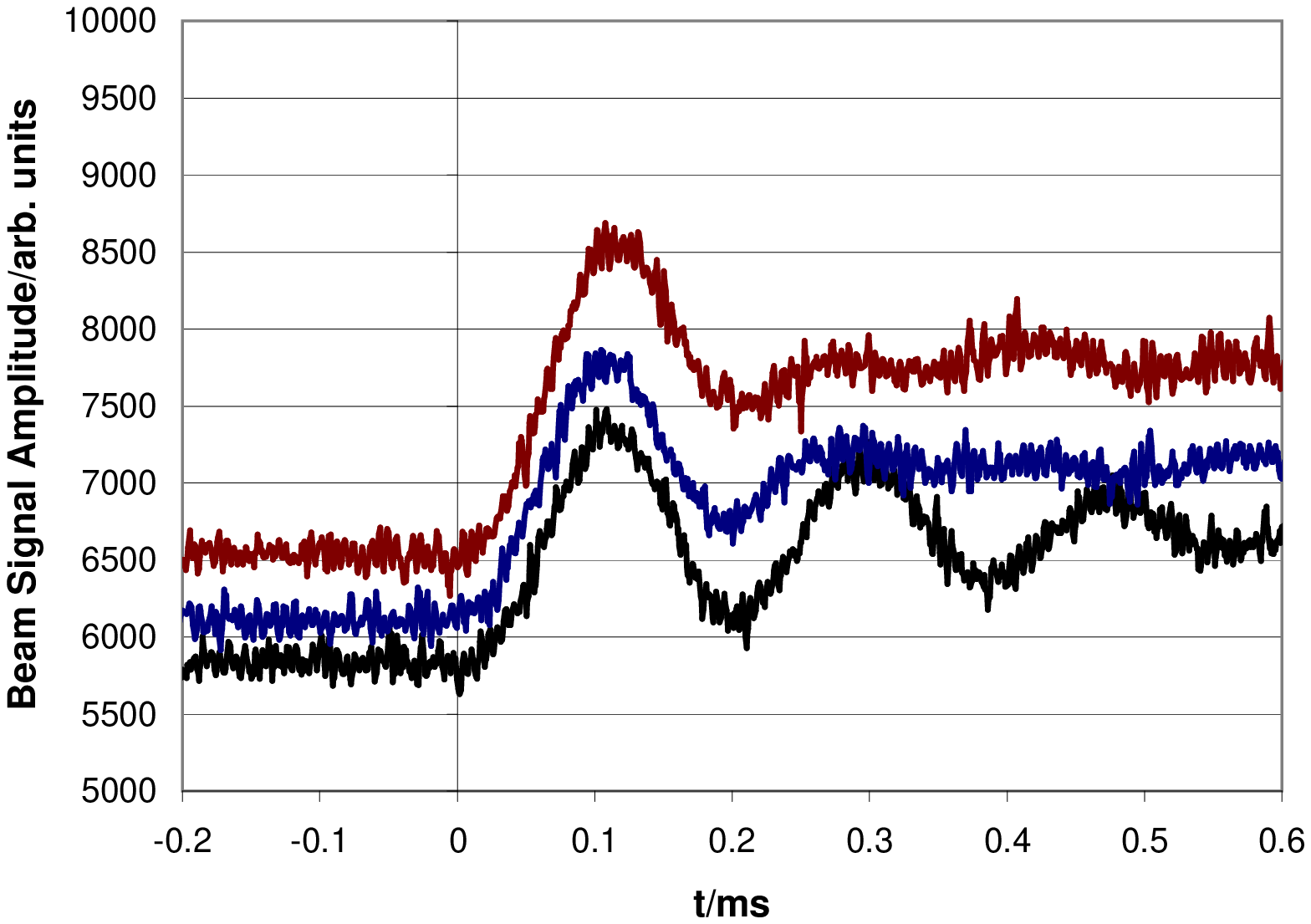,width=3in}
\psfrag{t}{$t/\rm ms$}
\psfrag{arbitraryunits}[c][c]{Beam Signal Amplitude/arb. units}
\epsfig{file=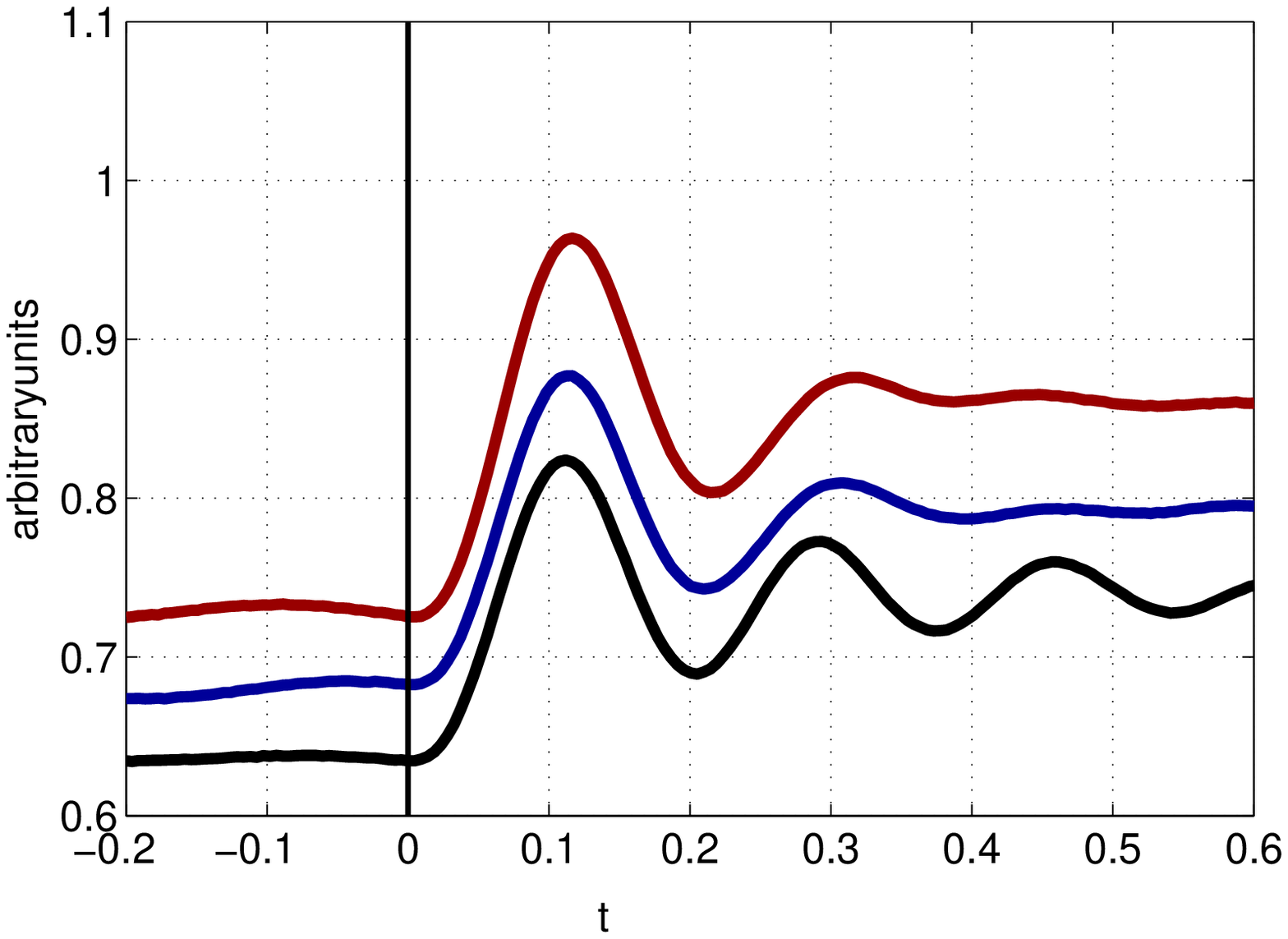,width=3in}
\caption{Measurement results (upper diagram) and 
simulation results (lower diagram) for beam signal amplitude 
(Uppermost trace: both control loops for quadrupole damping and for dipole damping on, 
trace in the middle: only control loop for quadrupole damping on, 
lowest trace: both control loops off)}
\label{figmeasurement}
\end{figure}

The upper diagram in Fig.~\ref{figmeasurement} shows the measured amplitude of the 
beam signal. The lowest trace is obtained if neither the beam phase control 
nor the quadrupole damping system are switched on. The oscillation frequency is
lower than $2 \; f_S$ (period of $200 \; \mu s$ instead of $150 \; \mu s$) since
the nonlinearity of the bucket cannot be neglected for the momentum spread of
the bunch. Due to Landau damping, the oscillation becomes weaker with time.   
The trace in the middle of Fig.~\ref{figmeasurement} 
shows that the damping time is reduced significantly in comparison with Landau 
damping if the quadrupole damping system is switched on. 
For the last uppermost trace in Fig.~\ref{figmeasurement} 
not only this quadrupole damping system was switched on, but also the 
beam phase control system which damps the coherent dipole mode. It is obvious
that both control loops work together without negative influence. 

\begin{figure}
\centering
\epsfig{file=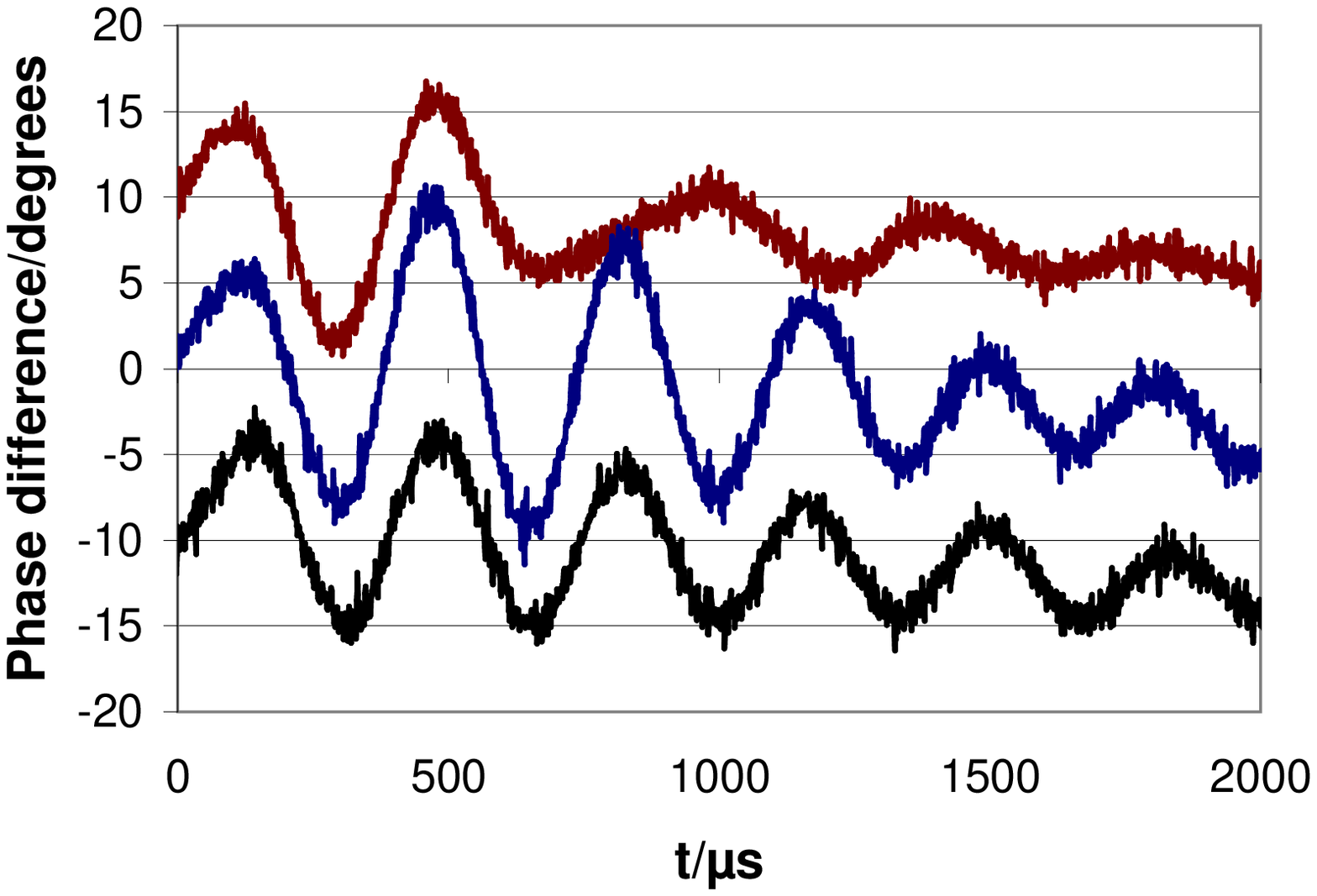,width=3in}
\psfrag{t}{$t/\rm \mu s$}
\psfrag{degree}[c][c]{Beam Signal Phase/degrees}
\epsfig{file=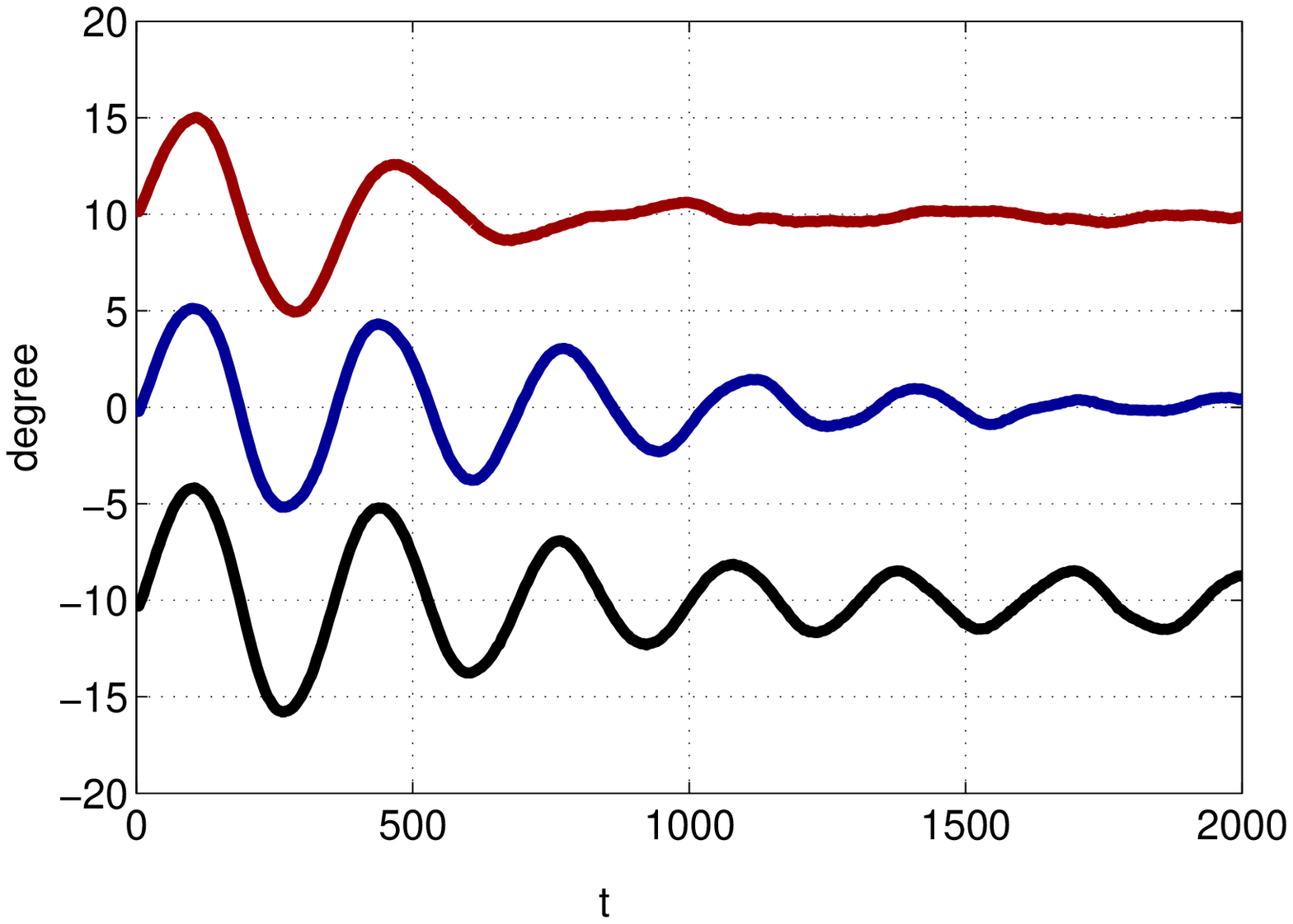,width=3in}
\caption{Measurement results (upper diagram) and 
simulation results (lower diagram) for beam signal phase 
(Uppermost trace: both control loops for quadrupole damping and for dipole damping on, 
trace in the middle: only control loop for quadrupole damping on, 
lowest trace: both control loops off)}
\label{figmeasdipole}
\end{figure}

The upper diagram in Fig.~\ref{figmeasdipole} shows the measured phase of the 
beam signal. 
A phase offset was introduced for each trace in order to have the same order 
as in Fig.~\ref{figmeasurement}. Therefore, the lowest trace again corresponds to 
both control loops being off which leads to Landau damping only. 
As expected, the oscillation period of $400 \; \rm \mu s$ equals twice the period 
of the quadrupole oscillation.    
The uppermost trace shows that the coherent dipole oscillation is damped 
faster than the Landau damping time if both control loops are switched on. 
The trace in the middle indicates that switching on the quadrupole damping system 
while disabling the dipole damping system increases the excitation of the 
coherent dipole mode (in comparison with the case that both loops are disabled).

\begin{figure}
\centering
\psfrag{t in msec}{t/ms}
\psfrag{y}{$y$}
\epsfig{file=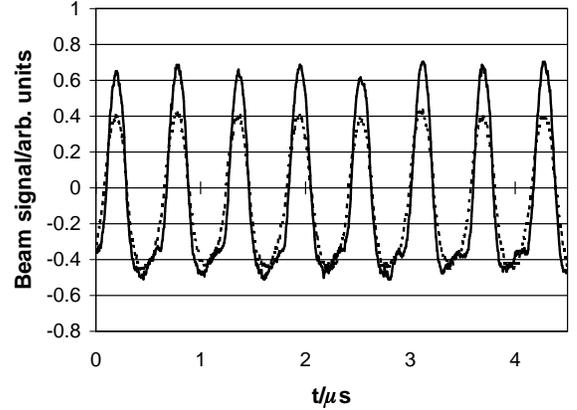,width=3in}
\caption{Measured beam signal 
(Dotted line: before voltage jump, solid line: first maximum after voltage jump)}
\label{figbeamsignal}
\end{figure}

\subsection{Simulation}
\label{simulation}

The experiments were compared with nonlinear tracking simulations. 
The simulation program used nonlinear discrete mapping equations in 
$\Delta \varphi$ and $\Delta W / \omega_{RF}$ for the 
longitudinal dynamics. The particle positions in phase space  were then 
converted to the $(x,y)$ plane with the variables $x=\Delta \varphi$ and 
$y= - \Delta \dot{\varphi}/\omega_S$.
The beam current signal was calculated as a histogram using bins on the 
$\Delta \varphi$-axis.
The beam signal amplitude and phase were obtained by an FFT of the 
beam current signal. 

Since coherent modes were excited in the experiment, only one bunch was 
simulated and compared with one bunch of the measured bunch signals at $h=8$. 
The macro-particles of the simulated bunch were initialized randomly with a
Gaussian probability distribution. The matched bunch was injected at the 
gap voltage of $5~\rm kV$. 
The gap voltage was modeled according to eqn.~(\ref{eqngapvoltage1})
where $\Delta \varphi_{gap}$ is the output of the dipole feedback loop 
and $\epsilon$ of the quadrupole feedback loop, respectively. 
The dipole feedback loop comprised an FIR filter and the dynamics of the 
DDS module which was modeled as an integrator (see Fig.~\ref{figblockdiagram} 
for the control loop topology). The input of the dipole 
feedback was the beam signal phase. The quadrupole feedback loop consisted 
of an FIR filter and an additional integral controller. 
The input of the quadrupole feedback was the absolute 
beam signal amplitude $A_1(t)$. 
The control loop parameters are given in Table \ref{tabloopparams}.

The cavity and its control loop dynamics 
were taken into account by a time constant of $T_{cav}=20~\rm \mu s$. 
In the experiment, the voltage step from $5~\rm kV$ to $10~\rm kV$ also causes 
a small phase shift due to the nonlinearity of the cavity system
(detuning effect). 
Therefore, this phase shift was also added to the simulation.

\subsection{Simulation Results and Comparison with Measurement}
\label{simresults}

In order to simulate the scenarios presented in section \ref{measresults}
using the procedure described in section \ref{simulation}, 
it was necessary to determine the beam parameters. For all three cases 
(only Landau damping without control loops, only quadrupole damping, 
quadrupole and dipole damping), the rms width of the beam signal 
before the voltage jump was determined by matching a Gaussian  
distribution to those measurement results which were valid before the 
voltage jump occurs. This fitting procedure leads to the  
parameters shown in Table \ref{tabvariances}. Since the bunch was matched
before the voltage jump occurred, we have $v_x=v_{x0}$ and $v_y=v_{y0}$.
 
\begin{table}
\caption{\label{tabvariances}Variances of measured beam before voltage step}
\begin{ruledtabular}
\begin{tabular}{llll}
& $v_x=v_{x0}$ & $v_y=v_{y0}$ & $v_0=\sqrt{v_x v_y}$\\ \hline
\textrm{only Landau} & 1.135 & 0.691 & 0.886\\
\textrm{damping} & & &  \\ \hline
\textrm{quadrupole} & 1.317 & 0.791 & 1.021\\ 
\textrm{damping on} & & & \\ \hline 
\textrm{dipole and quadrupole} & 1.313 & 0.792 & 1.020 \\
\textrm{damping on} &&&
\end{tabular}
\end{ruledtabular}
\end{table}

It is obvious that almost the same parameters are obtained for the three measurements 
which means that the beam quality was reproducible during the beam experiment.

The voltage step from $5~\rm kV$ to $10~\rm kV$ leads to an increase of the
bucket area by a factor of $\sqrt{2}$. Therefore, the relative bunch area $v_0$
will be $1/\sqrt{2}$ times smaller than before the step. 
The $(x,y)$ phase space was defined in such a way that the trajectories 
are circles. Therefore, $v_{x0}$ and $v_{y0}$ will also be 
$1/\sqrt{2}$ times smaller than before the step.  
In the case that no 
control loops are present (only Landau damping), this leads to 
$v_{x0}=1.135/\sqrt{2}=0.803$. 
This value which is valid for the new larger bucket immediately after 
the voltage step was therefore approximately assumed in Table \ref{tabparam}. 
Based on equations (\ref{eqnfseff}) and (\ref{eqndphieff}), 
it leads to $f_{S,eff}=0.8 f_S=2650 \; \rm Hz$ which is in agreement with the 
measurement. 

Now that the initial conditions for the beam were fixed, further parameters shown 
in Table \ref{tabloopparams} had to be adopted from the beam experiment. 
The only two parameters that were now still open were the proportional gain 
factors for the two control loops. 
As Figures \ref{figmeasurement} and \ref{figmeasdipole} show, the proper choice 
of these gain factors leads to a good agreement between measurement and simulation. 

\subsection{Geometrical Interpretation for Matched Bunches}

For the first case (only Landau damping) we found $v_{x0}=1.135$ before the 
voltage step occurs. According to equation (\ref{eqndphieff}) this leads to 
\[
x_{eff}=\Delta \varphi_{eff}=2.13. 
\]
The $(x,y)$ phase space leads to a bucket with the bucket length $2 x_{max}$
where $x_{max}=\pi$ and with the bucket height $2 y_{max}$ where $y_{max}=2$.
The bucket area equals $16$. If $x$ is increased from $0$ to $\pi$, the
ratio 
\[
\frac{y}{x}=\frac{\sqrt{2 (1-\cos \; x)}}{x}
\]
will decrease from $1$ (circular trajectory) to $2/\pi$ (separatrix). 
If we apply this formula for $x_{eff}$, we obtain $y_{eff}/x_{eff}=0.8214$. 
Due to equation (\ref{eqndphieff}) this leads to $v_{y0}/v_{x0}=0.67$. 
This is close to
the ratio $v_{y0}/v_{x0}=0.61$ derived from the data in Table \ref{tabvariances}. 
This is an indication that $x_{eff}=2 \sqrt{v_{x0}}$ and 
$y_{eff}=2 \sqrt{v_{y0}}$ may actually be interpreted as the effective half axes 
of the bunch. In our case this leads to a bucket filling factor of 
\[
\frac{\pi x_{eff} y_{eff}}{16}=
\frac{\pi}{16} x_{eff}^2 \frac{y_{eff}}{x_{eff}}=73 \; \% 
\]  
This is close to the value $\pi v_0/4=70 \; \%$ based on the values in 
Table \ref{tabvariances}.  

Instead of the fitting procedure applied above, 
the value $x_{eff}=\Delta \varphi_{eff}$ may also be found by determining the 
two-sigma length of the bunch (in our case, the one-sigma length derived from 
the measured bunch profile is about $0.1 \; \mu s$ which corresponds 
to $\Delta \varphi_\sigma=1.08$). Based on this value, one may determine 
$v_{x0}=\Delta \varphi_\sigma^2$, and $v_{y0}/v_{x0}=(y_{eff}/x_{eff})^2$.  
The ratio $f_{S,eff}/f_S$ results from eqn.~(\ref{eqnsyncfreq1}).
Hence, all relevant parameters required for the simulation can easily be
determined based on the bunch length. 

The accuracy of the formulas presented in the paper at hand could 
still be improved by modifying the theoretical factors, but it is sufficient 
for the basic understanding and for the control system design.   
 
\section{Conclusion}
Several specific models for longitudinal beam oscillations have been 
analyzed by different methods leading to the following observations: 
\begin{itemize}
\item The definitions of the longitudinal modes in 
phase space by the bunch shape or in the frequency domain by the 
spectral lines are not strictly equivalent in general.
\item In a strict sense, the term 'mode' usually implies linearity of the 
system. The examples in this paper show that both the dipole mode and the 
quadrupole may occur in a linear bucket whereas the sextupole mode requires 
a nonlinear bucket. 
\item It was shown that the mode $m=1$ shows primarily a phase modulation 
with the frequency $f_S$ whereas the mode $m=2$ shows primarily an amplitude
modulation with the frequency $2 f_S$. According to the ODE solutions, it is 
also possible to damp these oscillations using the same type of modulation. 
Due to equation (\ref{gleqnphigap1}), a gap voltage modulation additionally acts  
in the same way as a phase modulation of the gap voltage whereas the reverse is
not true.  
For higher orders $m>2$, we do not find phase or amplitude modulations in 
the linear model, and it is also impossible to excite a sextupole oscillation. 
Therefore, the case $m>2$ differs significantly from the case $m \le 2$
and must be analyzed using nonlinear models.
\item For control loop analysis, it is possible to use the quantities 
$\bar x$ and $v_x$ instead of phase and amplitude information. These quantities 
are easier to determine since no projection of phase space onto the time axis 
is required.  
\item The equations for $\bar{x}$ and $v_x$ allow an estimation of the area 
of stability for the dipole and quadrupole damping systems. 
This is not restricted to the FIR filters used here and could also be
applied to other control algorithms.
\item The connection between the longitudinal coherent modes,
the bunch mean and rms values, and the beam signal phase and amplitude signals 
has been demonstrated. 
\item There is reason to assume that the moment approach can be extended 
to the sextupole and higher order modes
thus enabling the controller design for the higher order modes.
\item The commonly used transfer functions for the dipole and 
quadrupole oscillation are only approximations in the scope of our model.
\end{itemize}
As a conclusion, the models described in the paper at hand allow a better 
understanding of longitudinal modes of oscillation and their damping by
active feedback systems.

\appendix

\section{Linearization}
\label{linearization}

The derived equations (\ref{eqnxbar1}), (\ref{eqnybar1}), (\ref{eqnddotvx1}),
and (\ref{glvydot2})
\begin{eqnarray*}
 \dot{\bar{x}} &=& - \omega_S \bar{y} \\
 \dot{\bar{y}} &=& \omega_S (1+\epsilon) (\bar{x}-\Delta \tilde{\varphi}_{gap}) \\
 \ddot{v}_x &=& 2\omega_S^2 v_y - 2 \omega_S^2 (1+\epsilon)v_x \\
 \dot{v}_y &=& -(1+\epsilon)\dot{v}_x,
\end{eqnarray*}
can be written as a state-space model
\[
 \dot{\bf x} = {\bf f}({\bf x},\epsilon,\tilde{\varphi}_{gap})
\]
with the state vector
\[
 \bf{x} = \begin{pmatrix} x_1 & x_2 & x_3 & x_4 & x_5 \end{pmatrix}^\mathrm{T} = 
\begin{pmatrix} \bar{x} & \bar{y} & v_x & \dot{v}_x & v_y \end{pmatrix}^\mathrm{T}
\]
and the nonlinear function
\[
 {\bf f}({\bf x},\epsilon,\Delta \tilde{\varphi}_{gap})
= \begin{pmatrix} - \omega_S x_2 \\ \omega_S(1+\epsilon)(x_1-\Delta \tilde{\varphi}_{gap}) \\
x_4 \\ 2 \omega_S^2 x_5 - 2 \omega_S^2 (1+\epsilon)x_3 \\ -(1+\epsilon)x_4 \end{pmatrix}.
\]
In the following, a linearization with $\Delta {\bf x}={\bf x}-{\bf x}_{op}$
around the operation point
\[
 {\bf x}_{op} = \begin{pmatrix} 0 & 0 & v_0 & 0 & v_0 \end{pmatrix}^\mathrm{T}, \quad
\epsilon_{op} = 0, \quad \Delta \tilde{\varphi}_{gap}=0
\]
is performed, which corresponds to the matched circle-shaped bunch. 
This linearization (cf.~\citep{SlotineLi}) leads to the linear system
\begin{align}
\label{eq:linearsystem}
 \Delta \dot{\bf x}(t) = {\bf A} \Delta {\bf x}(t) + {\bf b}_1 \epsilon(t) + 
 {\bf b}_2 \Delta \tilde{\varphi}_{gap}(t)
\end{align}
with the system matrix
\[
 {\bf A} = \left. \frac{\partial {\bf f}}{\partial \bf{x}} \right|_{op}
= \begin{pmatrix} 0 & -\omega_S & 0 & 0 & 0 \\
\omega_S & 0 & 0 & 0 & 0 \\
0 & 0 & 0 & 1 & 0 \\
0 & 0 & -2\omega_S^2 & 0 & 2 \omega_S^2 \\
0 & 0 & 0 & -1& 0 \end{pmatrix}
\]
and the input matrices
\[
 {\bf b}_1 = \left. \frac{\partial {\bf f}}{\partial \epsilon} \right|_{op}
= \begin{pmatrix} 0 & 0 & 0 & -2\omega_S^2 v_0 & 0 \end{pmatrix}^\mathrm{T}
\]
and
\[
 {\bf b}_2 = \left. \frac{\partial {\bf f}}{\partial \Delta \tilde{\varphi}_{gap}} \right|_{op}
= \begin{pmatrix} 0 & -\omega_S & 0 & 0 & 0 & 0 \end{pmatrix}^\mathrm{T}.
\]
Please note that $\bf{A}$ has a block diagonal structure with one block corresponding
to the dynamics of the bunch center $\bar{x}$ and one to the dynamics of the bunch
variance $v_x$. In addition, the bunch center is only influenced by 
$\Delta \tilde{\varphi}_{gap}$
and the bunch variance only by $\epsilon$. 

Comparing the equations for $\Delta x_3$ and $\Delta x_5$ in~\eqref{eq:linearsystem} 
yields $\Delta \dot{x}_3(t) = - \Delta \dot{x}_5(t)$
and thus
\begin{align}
\label{eq:cond_x3x5}
\Delta x_3(t) + \Delta x_5(t) = v_x(t) + v_y(t) - 2v_0 = const.
\end{align}
which implies that the bunch variances are connected by an algebraic equation and
cannot be controlled independently. It must in principle be possible 
that the solution $\bf x$ of the differential equation 
reaches the operation point ${\bf x}_{op}$ 
(e.g. as an initial condition). For the operation point,
\begin{equation}
\Delta x_3 + \Delta x_5=0
\label{eqninitial1}
\end{equation}
is valid which therefore holds in general for any $t$ due to 
equation (\ref{eq:cond_x3x5}). 

With~\eqref{eq:linearsystem} and~\eqref{eqninitial1}, linear differential equations
of second order can be derived for the bunch center using $\Delta x_1 = \bar{x}$
and for the bunch variance using $\Delta x_3 = v_x - v_0$:
\[
\ddot{\bar{x}} + \omega_S^2 \bar{x} = \omega_S^2 \Delta \tilde{\varphi}_{gap}
\]
\[
\ddot{v}_x + 4\omega_S^2 (v_x-v_0) = -2 \omega_S^2 v_0 \epsilon.
\]

\section{Fourier Series of the Dipole Oscillation Signal}
\label{fourier_series}

\begin{figure}
\centering
\epsfig{file=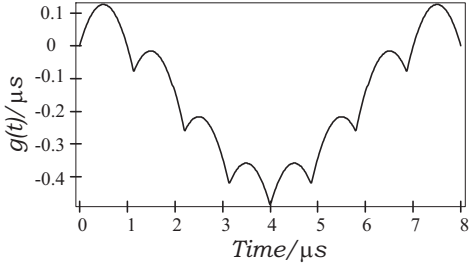,width=2.5in}
\caption{Visualization of the function $g(t)$ 
for $T_R=1 \;\mu s$, $T_S=8 \;\mu s$, $\Delta t=0.3 \; \mu s$}
\label{figdipolefourier}
\end{figure}

As an abbreviation, we set
\[
T_k=k T_R+\Delta t \; sin\left(2 \pi k \frac{T_R}{T_S} \right)
\mbox{\qquad} \Delta t<T_R/2
\]
and 
\[
a_k(t)=t-T_k. 
\]

We define one period of a special periodic function by
\[
g(t)=\frac{tT_S-t^2}{2 T_R}+
\sum_{k=1}^{\frac{T_S}{T_R}-1} \left[a_k(t) \sigma(a_k(t))-\frac{t}{T_S} a_k(T_S) \right]
\]
in the range $0 \le t \le T_S$ ($g(t)=0$ elsewhere) such that the 
periodic function becomes
\[
f^{(0)}(t)=\sum_{m=-\infty}^{+\infty} g(t-m T_S).
\]
$\sigma(t)$ denotes the unit step function. An example for $g(t)$ is 
shown in Fig.~\ref{figdipolefourier}.
This function $f^{(0)}(t)$ is a continuous even function
(note that $g(0)=g(T_S)=0$). According to  
\[
a^{(p)}_n=\frac{2}{T_S} \int_0^{T_S} f^{(p)}(t) \; cos(n \omega_S t) \; dt 
\]
\[
b^{(p)}_n=\frac{2}{T_S} \int_0^{T_S} f^{(p)}(t) \; sin(n \omega_S t) \; dt 
\]
we may derive the Fourier coefficients in order to get the series representation
\begin{equation}
f^{(p)}(t)=\frac{a^{(p)}_0}{2}+ 
\sum_{n=1}^\infty a^{(p)}_n cos(n \omega_S t)+
b^{(p)}_n sin(n \omega_S t).
\label{eqnfourier1}
\end{equation}
A lengthy but straightforward calculation leads to  
\[
a^{(0)}_n=-\frac{2}{T_S (n\omega_S)^2} 
\left[1+\sum_{k=1}^{\frac{T_S}{T_R}-1} cos \left(n \omega_S T_k \right)\right]
\mbox{\quad for } n \ne 0,
\]
\[
a^{(0)}_n=\frac{T_S^2}{6 T_R}+
\sum_{k=1}^{\frac{T_S}{T_R}-1} \left(\frac{T_k^2}{T_S}-T_k \right)
\mbox{\quad for } n=0,
\]
\[
b^{(0)}_n=0 \mbox{\quad for all } n.
\]

It is obvious that for $n \ne 0$, 
\[
|a^{(0)}_n| \le \frac{2}{T_R (n\omega_S)^2}
\]
holds. Hence, we have $|a^{(0)}_n|<M n^\kappa$ with real constants $M, \kappa$.  
According to \citep{Zemanian}, Corollary 2.4-3b, the series $f^{(p)}(t)$ 
in equation (\ref{eqnfourier1}) converges in the space $\cal D'$ of distributions.  
In $\cal D'$ it is allowed to differentiate this series without 
further requirements (\citep{Zemanian}, Corollary 2.4-3a). For the function
$g(t)$, we obtain 
\[
g'(t)=\frac{T_S-2t}{2 T_R}+
\sum_{k=1}^{\frac{T_S}{T_R}-1} \left[\sigma(a_k(t))-\frac{1}{T_S} a_k(T_S) \right]
\]
(for $0 < t < T_S$, $g'(t)=0$ elsewhere). 
We find 
\begin{eqnarray*}
g'(0+)&=& 1-\frac{T_S}{2T_R}+\sum_{k=1}^{\frac{T_S}{T_R}-1} \frac{T_k}{T_S}\\
g'(T_S-)&=& -\frac{T_S}{2T_R}+\sum_{k=1}^{\frac{T_S}{T_R}-1} \frac{T_k}{T_S}
\end{eqnarray*}
such that 
\[
f^{(1)}(t)=\sum_{m=-\infty}^{+\infty} g'(t-m T_S)
\]
(regarded as a locally integrable function) has unit steps at $t=mT_S$.
Both $g'$ and $f^{(1)}$ are locally integrable functions and may therefore be
regarded as regular distributions.  

As mentioned above, we may now differentiate this distribution. Thus 
we obtain in $\cal D'$: 
\[
f^{(2)}(t)=-\frac{1}{T_R}+\sum_{k=-\infty}^{+\infty} \delta(a_k(t))
\]
On the other hand, the derivative of eqn.~(\ref{eqnfourier1}) yields 
\begin{eqnarray*}
f^{(2)}(t)&=&\sum_{n=1}^\infty a^{(2)}_n cos(n \omega_S t)=\\
&=&-\sum_{n=1}^\infty a^{(0)}_n (n \omega_S)^2 cos(n \omega_S t)
\end{eqnarray*}
with 
\[
a^{(2)}_n=\frac{2}{T_S} 
\left[1+\sum_{k=1}^{\frac{T_S}{T_R}-1} cos \left(n \omega_S T_k \right)\right]
\mbox{\quad for } n \ne 0.
\]
This completes the proof of eqn.~(\ref{eqndirac1}).

\begin{acknowledgments}
The authors would like to thank 
Professor Dr.~J\"{u}rgen Adamy,  
Priv.-Doz.~Dr.~habil.~Peter H\"{u}lsmann, 
Dr.~Roland Kempf, 
Dr.~Ulrich Laier, and 
Dr.~Gerald Schreiber  
for many fruitful discussions.
Parts of the work were carried out in the scope of the project SIS100, task  
'Longitudinal Feedback System' in the EU FP6 Design Program. 
Another part was funded by the VW-Stiftung and the Deutsche Telekom Stiftung.
\end{acknowledgments}

\bibliographystyle{plainnat} 
\bibliography{beam_osc}

\providecommand{\noopsort}[1]{}\providecommand{\singleletter}[1]{#1}%
\begin{thebibliography}{10}%
\makeatletter
\providecommand \@ifxundefined [1]{%
 \@ifx{#1\undefined}
}%
\providecommand \@ifnum [1]{%
 \ifnum #1\expandafter \@firstoftwo
 \else \expandafter \@secondoftwo
 \fi
}%
\providecommand \@ifx [1]{%
 \ifx #1\expandafter \@firstoftwo
 \else \expandafter \@secondoftwo
 \fi
}%
\providecommand \natexlab [1]{#1}%
\providecommand \enquote  [1]{``#1''}%
\providecommand \bibnamefont  [1]{#1}%
\providecommand \bibfnamefont [1]{#1}%
\providecommand \citenamefont [1]{#1}%
\providecommand \href@noop [0]{\@secondoftwo}%
\providecommand \href [0]{\begingroup \@sanitize@url \@href}%
\providecommand \@href[1]{\@@startlink{#1}\@@href}%
\providecommand \@@href[1]{\endgroup#1\@@endlink}%
\providecommand \@sanitize@url [0]{\catcode `\\12\catcode `\$12\catcode
  `\&12\catcode `\#12\catcode `\^12\catcode `\_12\catcode `\%12\relax}%
\providecommand \@@startlink[1]{}%
\providecommand \@@endlink[0]{}%
\providecommand \url  [0]{\begingroup\@sanitize@url \@url }%
\providecommand \@url [1]{\endgroup\@href {#1}{\urlprefix }}%
\providecommand \urlprefix  [0]{URL }%
\providecommand \Eprint [0]{\href }%
\@ifxundefined \urlstyle {%
  \providecommand \doi  [0]{\begingroup \@sanitize@url \@doi}%
  \providecommand \@doi [1]{\endgroup \@@startlink {\doibase
  #1}doi:\discretionary {}{}{}#1\@@endlink }%
}{%
  \providecommand \doi  [0]{doi:\discretionary{}{}{}\begingroup
  \urlstyle{rm}\Url }%
}%
\providecommand \doibase [0]{http://dx.doi.org/}%
\providecommand \Doi [0]{\begingroup \@sanitize@url \@Doi }%
\providecommand \@Doi  [1]{\endgroup\@@startlink{\doibase#1}\@@Doi}%
\providecommand \@@Doi [1]{#1\@@endlink}%
\providecommand \selectlanguage [0]{\@gobble}%
\providecommand \bibinfo  [0]{\@secondoftwo}%
\providecommand \bibfield  [0]{\@secondoftwo}%
\providecommand \translation [1]{[#1]}%
\providecommand \BibitemOpen [0]{}%
\providecommand \bibitemStop [0]{}%
\providecommand \bibitemNoStop [0]{.\EOS\space}%
\providecommand \EOS [0]{\spacefactor3000\relax}%
\providecommand \BibitemShut  [1]{\csname bibitem#1\endcsname}%
\bibitem [{\citenamefont {Sacherer}(1973)}]{Sacherer1973}%
  \BibitemOpen
  \bibfield  {author} {\bibinfo {author} {\bibfnamefont {F.~J.}\ \bibnamefont
  {Sacherer}},\ }in\ \href@noop {} {\emph {\bibinfo {booktitle} {Proc. 5th IEEE
  Particle Accelerator Conference}}}\ (\bibinfo {organization} {IEEE},\
  \bibinfo {address} {San Francisco},\ \bibinfo {year} {1973})\ pp.\ \bibinfo
  {pages} {825--829}\BibitemShut {NoStop}%
\bibitem [{\citenamefont {Pedersen}\ and\ \citenamefont
  {Sacherer}(1977)}]{Pedersen1977}%
  \BibitemOpen
  \bibfield  {author} {\bibinfo {author} {\bibfnamefont {F.}~\bibnamefont
  {Pedersen}}\ and\ \bibinfo {author} {\bibfnamefont {F.}~\bibnamefont
  {Sacherer}},\ }\href@noop {} {\bibfield  {journal} {\bibinfo  {journal} {IEEE
  Trans. Nucl. Sci.},\ }\textbf {\bibinfo {volume} {24}},\ \bibinfo {pages}
  {1396} (\bibinfo {year} {1977})}\BibitemShut {NoStop}%
\bibitem [{\citenamefont {Lee}(1999)}]{lee}%
  \BibitemOpen
  \bibfield  {author} {\bibinfo {author} {\bibfnamefont {S.~Y.}\ \bibnamefont
  {Lee}},\ }\href@noop {} {\emph {\bibinfo {title} {Accelerator Physics}}}\
  (\bibinfo  {publisher} {World Scientific},\ \bibinfo {address} {Singapore},\
  \bibinfo {year} {1999})\BibitemShut {NoStop}%
\bibitem [{\citenamefont {Kamke}(1956)}]{Kamke}%
  \BibitemOpen
  \bibfield  {author} {\bibinfo {author} {\bibfnamefont {E.}~\bibnamefont
  {Kamke}},\ }\href@noop {} {\emph {\bibinfo {title} {Differentialgleichungen -
  L\"osungsmethoden und L\"osungen}}}\ (\bibinfo  {publisher} {Akademische
  Verlagsgesellschaft Geest \& Portig K.-G.},\ \bibinfo {address} {Leipzig},\
  \bibinfo {year} {1956})\BibitemShut {NoStop}%
\bibitem [{\citenamefont {MacLachlan}(1997)}]{MacLachlan1997}%
  \BibitemOpen
  \bibfield  {author} {\bibinfo {author} {\bibfnamefont {J.~A.}\ \bibnamefont
  {MacLachlan}},\ }in\ \href@noop {} {\emph {\bibinfo {booktitle} {Proc. 17th
  IEEE Particle Accelerator Conference}}}\ (\bibinfo {organization} {IEEE},\
  \bibinfo {address} {Vancouver},\ \bibinfo {year} {1997})\ pp.\ \bibinfo
  {pages} {2556--2558}\BibitemShut {NoStop}%
\bibitem [{\citenamefont {Gross}(2009)}]{Gross}%
  \BibitemOpen
  \bibfield  {author} {\bibinfo {author} {\bibfnamefont {K.}~\bibnamefont
  {Gross}},\ }\emph {\bibinfo {title} {Regelung koh{\"a}renter longitudinaler
  Schwingungen eines gebunchten Strahls in einem Schwerionensynchrotron}},\
  \href@noop {} {\bibinfo {type} {Diploma thesis}},\ \bibinfo  {school}
  {Technical University Darmstadt} (\bibinfo {year} {2009})\BibitemShut
  {NoStop}%
\bibitem [{\citenamefont {Boussard}(1991)}]{Boussard1991}%
  \BibitemOpen
  \bibfield  {author} {\bibinfo {author} {\bibfnamefont {D.}~\bibnamefont
  {Boussard}},\ }\href@noop {} {\emph {\bibinfo {title} {Design of a Ring RF
  System}}},\ \bibinfo {type} {Internal Report SL/91-2 (RFS)}\ (\bibinfo
  {institution} {CERN},\ \bibinfo {year} {1991})\BibitemShut {NoStop}%
\bibitem [{\citenamefont {Klingbeil}\ \emph {et~al.}(2007)\citenamefont
  {Klingbeil}, \citenamefont {Zipfel}, \citenamefont {Kumm},\ and\
  \citenamefont {Moritz}}]{Klingbeil2007}%
  \BibitemOpen
  \bibfield  {author} {\bibinfo {author} {\bibfnamefont {H.}~\bibnamefont
  {Klingbeil}}, \bibinfo {author} {\bibfnamefont {B.}~\bibnamefont {Zipfel}},
  \bibinfo {author} {\bibfnamefont {M.}~\bibnamefont {Kumm}}, \ and\ \bibinfo
  {author} {\bibfnamefont {P.}~\bibnamefont {Moritz}},\ }\href@noop {}
  {\bibfield  {journal} {\bibinfo  {journal} {IEEE Trans. Nucl. Sci.},\
  }\textbf {\bibinfo {volume} {54}},\ \bibinfo {pages} {2604} (\bibinfo {year}
  {2007})}\BibitemShut {NoStop}%
\bibitem [{\citenamefont {Slotine}\ and\ \citenamefont {Li}(1991)}]{SlotineLi}%
  \BibitemOpen
  \bibfield  {author} {\bibinfo {author} {\bibfnamefont {J.-J.~E.}\
  \bibnamefont {Slotine}}\ and\ \bibinfo {author} {\bibfnamefont
  {W.}~\bibnamefont {Li}},\ }\href@noop {} {\emph {\bibinfo {title} {Applied
  Nonlinear Control}}}\ (\bibinfo  {publisher} {Prentice Hall},\ \bibinfo
  {address} {Englewood Cliffs, New Jersey},\ \bibinfo {year}
  {1991})\BibitemShut {NoStop}%
\bibitem [{\citenamefont {Zemanian}(1987)}]{Zemanian}%
  \BibitemOpen
  \bibfield  {author} {\bibinfo {author} {\bibfnamefont {A.~H.}\ \bibnamefont
  {Zemanian}},\ }\href@noop {} {\emph {\bibinfo {title} {Distribution Theory
  and Transform Analysis}}}\ (\bibinfo  {publisher} {Dover Publications,
  Inc.},\ \bibinfo {address} {New York},\ \bibinfo {year} {1987})\BibitemShut
  {NoStop}%
\end{thebibliography}%

\end{document}